\documentclass[%
 reprint,
nofootinbib,
 amsmath,amssymb,
 aps,
pra,
]{revtex4-2}

\usepackage{graphicx}
\usepackage{dcolumn}
\usepackage{bm}
\usepackage{physics}
\usepackage[dvipsnames]{xcolor}
\usepackage{comment}
\usepackage{amsthm}
\usepackage{mathrsfs}
\usepackage{float}
\usepackage[colorlinks=true,bookmarks=false,citecolor=blue,urlcolor=blue,linkcolor=blue]{hyperref}
\definecolor{lublue}{RGB}{0, 100, 255}

\begin{document}

\newcommand{\ie}{\emph{i.e.}}
\newcommand{\eg}{\emph{e.g.}}
\newcommand{\cf}{\emph{cf.}}
\newcommand{\ea}{\emph{et al.}}
\newcommand{\etc}{\emph{etc.}}

\preprint{APS/123-QED}

\title{Optimized Laser Models with Heisenberg-Limited Coherence and Sub-Poissonian Beam Photon Statistics}

\author{L.~A.~Ostrowski}
 \email{lucas.ostrowski@griffithuni.edu.au}
\author{T.~J.~Baker}
\author{S.~N.~Saadatmand}
\author{H.~M.~Wiseman}
 \email{h.wiseman@griffith.edu.au}
\affiliation{%
 Centre for Quantum Dynamics, Griffith University, Yuggera Country, Brisbane, Queensland 4111, Australia
}%

\date{\today}

\begin{abstract}
Recently it has been shown that it is possible for a laser to produce a stationary beam with a coherence (quantified as the mean photon number at spectral peak) which scales as the fourth power of the mean number of excitations stored within the laser, this being quadratically larger than the standard or Schawlow-Townes limit~\cite{HL}. Moreover, this was analytically proven to be the ultimate quantum limit (Heisenberg limit) scaling under defining conditions for CW lasers, plus a strong assumption about the properties of the output beam. In Ref.~\cite{Ostrowski}, we show that the latter can be replaced by a weaker assumption, which allows for highly sub-Poissonian output beams, without changing the upper bound scaling or its achievability. In this Paper, we provide details of the calculations in Ref.~\cite{Ostrowski}, and introduce three new families of laser models which may be considered as generalizations of those presented in that work. Each of these families of laser models is parameterized by a real number, $p$, with $p=4$ corresponding to the original models. The parameter space of these laser families is numerically investigated in detail, where we explore the influence of these parameters on both the coherence and photon statistics of the laser beams. Two distinct regimes for the coherence may be identified based on the choice of $p$, where for $p>3$, each family of models exhibits Heisenberg-limited beam coherence, while for $p<3$, the Heisenberg limit is no longer attained. Moreover, in the former regime, we derive formulae for the beam coherence of each of these three laser families which agree with the numerics. We find that the optimal parameter is in fact $p\approx4.15$, not $p=4$.

\end{abstract}

\maketitle


\section{Introduction}

As the frontier of quantum technology enters the NISQ era, remarkable levels of control over quantum systems have been demonstrated~\cite{Chou, Sayrin}, bringing us to the point at which the properties of quantum mechanics are exploited to achieve feats that would be impractical by any classical device~\cite{McCormic, Gilmore, google, pan1, pan2, pan3, pan4}. However, in striving towards the quantum technologies of the future and the powerful applications they potentially offer, it remains imperative that quantum scientists and engineers alike continue pushing the envelope of discovery towards the enhanced control over quantum systems~\cite{Preskill}.

Recent work regarding the limit to the amount of optical coherence that can be produced by a laser marks an example of this~\cite{HL}. Coherence is a quantity that is of fundamental importance not only in the field of quantum technology~\cite{wineland, Bloch}, but to precision technology in general~\cite{Hecht}. In that work, the authors demonstrated that a \textit{quantum enhancement} is possible for the coherence, denoted by $\mathfrak{C}$, of a continuous wave (CW) laser. In this context, a quantum enhancement is said to exist if the ultimate limit imposed by quantum theory, or \textit{Heisenberg limit}, for the performance of a device scales better in terms of a particular resource when compared to the \textit{standard quantum limit} (SQL). 

The principled definition in Ref.~\cite{HL} of coherence---the mean photon number of the maximally populated mode of a field---implies that it is proportional to the product of the photon flux, $\mathcal{N}$, and the coherence time, $1/\ell$, for ``ideal" laser beams.. The authors of~\cite{HL} then derived the Heisenberg limit for a laser with such `ideal' properties (see below), $\mathfrak{C}_{\rm HL}^{\rm ideal} = \Theta(\mu^4)$. Here, $\mu$ is the mean number of photons within the laser, and is a critical resource for the production of a highly coherent beam of light. Thus, a quadratic enhancement over the Schawlow-Townes limit~\cite{ST}, $\mathfrak{C}_{\rm STL}^{\rm ideal} = \Theta(\mu^2)$, was theoretically demonstrated, forcing one to adopt the notion that this historic limit is only a SQL, and could be far surpassed by the Heisenberg limit. Note that throughout this Paper we make use of Bachmann-Landau notation to describe the limiting behaviour of particular functions~\cite{Burgisser}. The $\Theta$-notation given above means that $\mathfrak{C}_{\rm HL}^{\rm ideal}$, for instance, is bounded both above and below by $\mu^4$ asymptotically. Formally, this is to say that there exist two positive constants, $k_1$ and $k_2$, and a $\mu_0$, such that, for all $\mu>\mu_0$, we have $k_1\mu^4\leq|\mathfrak{C}_{\rm HL}^{\rm ideal}(\mu)|\leq k_2\mu^4$.

A model for a laser that produces a beam with Heisenberg-limited scaling of $\mathfrak{C}$ was also put forward in that work. The key element in that model is that both the input (gain) and output coupling are highly nonlinear. A potential implementation of such couplings on the platform of circuit-QED was given~\cite{HL}. Parallel to this, an independent group also demonstrated theoretically that the SQL for $\mathfrak{C}$ can be surpassed via a different proposal with a circuit-QED architecture~\cite{pekker}. Like the former model, the key element in the design was that the input and output coupling of the laser to its environment contains a nonlinear component. The circuit proposed in Ref.~\cite{pekker} may be more practical than that in Ref.~\cite{HL}, but, while it surpasses the SQL, with $\mathfrak{C}=\Theta(\mu^3)$, it does not achieve the Heisenberg limit.

In light of these recent theoretical results, the general focus of this Paper, along with our companion Letter~\cite{Ostrowski}, is to advance the understanding of lasers that perform at these extreme limits. We make the upper bound for $\mathfrak{C}$ more robust, where we rigorously demonstrate that $\mathfrak{C}=\Theta(\mu^4)$ remains the Heisenberg limit under much more relaxed conditions on the beam. This generalization has two motivations, as follows. 

First, it is quite likely that the most feasible laser model to implement on near term experimental hardware which surpasses the SQL for $\mathfrak{C}$ would not exactly satisfy the strict criterion placed on the beam in Ref.~\cite{HL}. This is because, unlike standard lasers, there is no intrinsic reason that engineered systems which produce a Heisenberg-limited beam should exhibit Poissonian beam photon statistics, for example. In particular, to have a Poissonian beam for a Heisenberg-limited laser requires specific engineering of the gain and loss processes~\cite{Ostrowski}. Because the implementation of these processes will be less-than-perfect in any experimental system, deviations in the photon statistics are to be expected. It is therefore important to consider bounds on the coherence for a broader class of beam statistics that would apply in such experiments.

Second, with this more general Heisenberg limit, we are able to provide a fundamental insight to the nature of laser radiation. This is because it does not rule out beams that can be highly sub-Poissonian~\cite{Davidovich}, and therefore permits a study into nonclassical photon statistics of lasers which perform at these extreme limits. Sub-Poissonian light generated from laser devices has been a topic of interest among quantum optics communities for decades due its potential applications in precision technology, including quantum-enhanced measurement, sensing, communication and information processing~\cite{Davidovich,Kolobov,Yamamoto1986b,Polzik,Ralph1995,Golubeva2008,Walls_Milb,Korolev2019,Mork2020,Goldberg,Hosseinidehaj2022,Zhao2022}. Early work on sub-Poissonian light generation from lasers was pioneered by the theoretical work of Golubev and Sokolov~\cite{Golubev1984}, who showed that this can be achieved by regularizing the pumping of excitations into the laser system. This was demonstrated in experiments shortly thereafter, conducted by Machida \textit{et al}.~\cite{Machida1987}, and Richardson and Shelby~\cite{Richardson1990} with semiconductor lasers. More recent interest in the production of sub-Poissonian laser light has centred on utilizing novel gain attributes made available on cavity- and circuit-QED platforms~\cite{Choi2006,Koppenhofer2016,Koppenhofer2017,Canela}.

Our particular focus here is the question of whether a trade-off exists between coherence and the degree of sub-Poissonianity in a \textit{Heisenberg-limited} laser beam, two properties that are of predominant importance in precision technology. One might expect that such a tradeoff exists based on a rough intuition that the former property is greatest when phase noise is minimized, while the latter is greatest when the intensity noise is minimized; in a single mode field there is certainly a tradeoff between phase and intensity variance because of the uncertainty principle~\cite{BarnettPegg,PeggBarnett}. However, this single mode argument does not straight-forwardly generalize to a laser beam. Older studies addressed this question for lasers limited by the SQL for coherence. It was shown that for laser models with a linear output coupling (which necessarily achieve $\mathfrak{C}=\Theta(\mu^2)$ at best), it is possible to modify the pumping such that a sub-Poissonian output is achieved without significantly changing the rate of phase diffusion~\cite{Bergou_Phase,Benkert_Phase,wiseman1993,Ralph2004}. That is, there is no tradeoff between coherence and sub-Poissonianity for these SQL-coherence laser models (note that in other standard laser models, where a sub-Poissonian beam is instead generated through nonlinear absorption processes, a tradeoff between these two quantities \textit{does} exist~\cite{Wiseman1991}). In this work, we aim to generalize these results for Heisenberg-limited lasers; because these lasers operate with a vastly smaller phase diffusion rate to begin with, this fundamental question of whether a tradeoff exists is not obvious.

Answering this query is a primary goal of our companion Letter~\cite{Ostrowski}, where two families of laser models where developed that exhibit both Heisenberg-limited coherence and sub-Poissonian beam photon statistics (quantified by the Mandel-$Q$ parameter) for specific values of parameters. There, it is shown that there is a perfect correlation between an increase in the coherence and a decrease in the Mandel-$Q$ parameter of the beam within those families of models. Moreover, the maximum coherence attained in those families is significantly larger than that of the original model presented in Ref.~\cite{HL}. This demonstrates the opposite of a tradeoff: there is a ``win-win" situation between coherence and sub-Poissonianity in a Heisenberg-limited laser beam. That is, it appears to be advantageous for the optimization of the coherence if measures are taken to reduce the number fluctuations in the beam. 

As well as fleshing out many of the proofs and results in the Letter~\cite{Ostrowski}, we present a host of additional results for three families of laser models that exhibit Heisenberg-limited coherence. We provide a detailed exploration of the parameter spaces characterizing these families, as well as a formula for the coherence of the laser beams that agrees with numerical calculations. From this, we are able to gain physical insight into the compatible relationship between coherence and sub-Poissonianity for some of these laser models. We note here that all of these models could, \textit{in principle}, be realised experimentally, following the method of Ref.~\cite{HL}. However, the primary motivation for their development here is to explore the aforementioned tradeoff question.

The three families we consider may be regarded as generalizations of those given in Ref.~\cite{Ostrowski} and are conceived by making modifications to the gain and loss mechanisms between the laser and its surrounding environment from the original laser model of Ref.~\cite{HL}. Each can be aptly distinguished from one another by considering only their pumping mechanisms: The first involves a \textit{randomly pumped (Markovian), quasi-isometric gain}; the second involves a \textit{randomly pumped, non-isometric gain}; and the third involves a \textit{regularly pumped (non-Markovian), quasi-isometric gain}. For ease of reference, these will be referred to according to the key parameters that characterize them, being the $p$-\textit{family}, the $\lambda,p$-\textit{family} and the $q,p$-\textit{family} of Heisenberg-limited laser models, respectively. Both of the last two of these are shown to exhibit sub-Poissonian beam photon statistics, and both reduce to the first family in the Poissonian limit.

An important new aspect of the three families of laser models is the parameter $p$. This controls the variance of the steady-state laser cavity distribution and is shown to have a strong influence on the coherence, as two distinct regimes may be identified for each family with respect to this parameter. For values $p\gtrsim3$, we find the scaling of the coherence for all families is $\mathfrak{C}=\Theta(\mu^4)$. That is to say that $\mathfrak{C}$ is Heisenberg-limited within this regime. Moreover, we find that a peak in the coherence in this regime is found at $p\approx4.15$ for each family, slightly different from the value $p=4$ used in Refs.~\cite{HL,Ostrowski}. On the other hand, for values $p\lesssim3$ (which means even broader cavity distributions), we find a change in exponent of the power law for the coherence and Heisenberg-limited scaling is lost. Here, we instead find the relationship between $\mathfrak{C}$ and $\mu$ to be approximately $\mathfrak{C}=\Theta(\mu^{p+1})$. We were able to reproduce all of this behaviour by heuristic arguments.

The remainder of this Paper is structured as follows. In Section~\ref{section2} we introduce and discuss the key concepts and quantities of interest throughout this Paper: the coherence and sub-Poissonianity of a laser beam, and our adopted measures of them. The conditions we place on a laser and its beam to derive the Heisenberg limit for the coherence are also summarized here, and we outline the differences between these and those given in Ref.~\cite{HL}. In Section~\ref{section3} we derive an upper bound for $\mathfrak{C}$ under these four conditions, which allows for beams that can be highly sub-Poissonian. In Section~\ref{section4} we briefly review the basic description of a laser in an iMPS framework and the numerical methods employed to compute the physical quantities of interest within this framework. In Section~\ref{V} we briefly provide some comments on practicality for the implementation of Heisenberg-limited laser dynamics, and provide some further justification to our claim regarding importance of considering the beam photon statistics for these efforts. In Section~\ref{Families} we introduce the three families of laser models which will be the subject of analysis for the subsequent sections of the Paper. Section~\ref{Numerical_Analysis} provides a numerical analysis of these three families of laser models, where we explore the parameter spaces characterizing each family in detail and identify interesting behaviour of the physical quantities which quantify the beam coherence and degree of beam sub-Poissonianity. In Section~\ref{general_p} we provide an analysis of the coherence, where we are able to derive formulae for our three families of models in the ``Heisenberg-limited regime" characterized by $p\gtrsim3$. We are able to show that these formulae accurately reproduce our numerical results given in Section~\ref{Numerical_Analysis} and based on this analysis, we are able to provide arguments as to why Heisenberg-limited coherence is lost for $p\lesssim3$. Finally, in Section~\ref{summary} we discuss our results and provide some concluding remarks.

\section{Lasers, Coherence and Sub-Poissonianity}
\label{section2}

\subsection{Laser Coherence}

Consider a one-dimensional bosonic beam travelling at fixed speed with translationally invariant statistics, which can be described by the single-parameter field operator $\hat{b}(t)$. Our adopted measure of coherence for such a beam may be defined generally as the mean number of photons in the maximally populated spatial mode~\cite{HL}. This amounts to the mathematical statement
\begin{align}\label{coh_general}
    \mathfrak{C} := \max_{u\in\mathfrak{u}}\langle \hat{b}_u^\dagger \hat{b}_u \rangle,
\end{align}
where $\hat{b}_u = (1/\sqrt{I_u})\int_{-\infty}^\infty dt u(t)\hat{b}(t)$ defines the annihilation operator for mode $u$. Here, $I_u = \int_{-\infty}^\infty dt|u(t)|^2$ is a normalization factor, while the maximization is over the modes $\mathfrak{u}$ within a particular frequency band; this avoids the trouble associated with a thermal state with arbitrarily low frequency having an exceedingly large coherence. 

For a beam with translationally invariant statistics, the mode $u$ which attains the maximum in Eq.~(\ref{coh_general}) is characterised by a flat waveform. Strictly, such a mode is not in $\mathfrak{u}$ as it is not square integrable, but we can consider $u_T(t) \propto e^{-|t|/T}$ and afterwards take the limit $T\xrightarrow{}\infty$. Consequently, $\mathfrak{C}$ is directly proportional to the maximum of the power spectrum~\cite{HL}, and as it is possible to redefine $\hat{b}(t)$ so as to absorb the rotation at the spectral peak frequency, the coherence may be expressed as
\begin{align}
    \mathfrak{C} = \int_{-\infty}^\infty ds G^{(1)}(t,s).
\end{align}
Here the $n^{\rm th}$-order Glauber coherence functions are defined in the usual way~\cite{Glauber}
\begin{align}\label{n_corrs}
    G^{(n)}(s_1,...,s_{2n}) := \langle \hat{b}^\dagger(s_1) \dotsm \hat{b}^\dagger(s_n)\hat{b}(s_{n+1})\dotsm\hat{b}(s_{2n}) \rangle,
\end{align}
with corresponding normalized forms
\begin{align}\label{n_corrs_norm}
    g^{(n)}(s_1,...,s_{2n}) := \frac{\langle \hat{b}^\dagger(s_1) \dotsm \hat{b}^\dagger(s_n)\hat{b}(s_{n+1})\dotsm\hat{b}(s_{2n}) \rangle}{\Pi_{i=1}^{2n}|G^{(1)}(s_i,s_i)|^{1/2}}.
\end{align}
With translationally invariant statistics, the denominator of Eq.~(\ref{n_corrs_norm}) can be expressed as $\Pi_{i=1}^{2n}|G^{(1)}(s_i,s_i)|^{1/2} = \mathcal{N}^n$, where we define
\begin{align}\label{flux}
    \mathcal{N} := G^{(1)}(t,t).
\end{align}
This quantity is interpreted as the photon flux from the laser.

It is well-established that, far above threshold, the state produced by an ideal standard CW laser in the absence of any technical noise can be described by a coherent state undergoing pure phase diffusion~\cite{Louisell, SSL, Carmichael}. That is, the beam can be described by an eigenstate 
\begin{align}\label{las_ideal}
    \ket{\beta(t)} = \ket{\sqrt{\mathcal N}e^{i\sqrt{\ell}W(t)}},
\end{align}
of $\hat{b}(t)$, where $W(t)$ represents a Wiener process. For such beams, the coherence may be evaluated in terms of $\mathcal{N}$ and the full-width at half-maximum (FWHM) of its Lorentzian Power spectrum (or linewidth), $\ell$, as
\begin{align}\label{coh_intuitive}
    \mathfrak{C} = \frac{4\mathcal{N}}{\ell}.
\end{align}
In this instance, we see that $\mathfrak{C}$ is has a straightforward interpretation, as roughly the number of photons that are emitted into the beam that are mutually coherent.

A noteworthy point regarding this measure of coherence is that it does not require an absolute phase or mean field, but is instead a statement of the beam's \textit{relative coherence}, and can be thought of the ability of the beam to act as a \textit{classical phase reference}. By considering the action on the field imposed by a beamsplitter~\cite{HL}, this coherence measure may be shown to be a monotonic function of the relative entropy of asymmetry~\cite{Vaccaro_asym}, which is a well established coherence measure employed in quantum information theory~\cite{quant_coherence, coherence_review}.

\subsection{Beam Photon Statistics}

As we have already alluded to, a key concept within this Paper is with regard to sub-Poissonian light in the \textit{output} field. The degree to which the output field is sub-Poissonian may be quantified by the Mandel-$Q$ parameter defined over the time duration $T$~\cite{Mandel1, Mandel2},
\begin{align}\label{mandel}
    Q_{T;T_0} := \frac{\langle (\Delta \hat{n}_{T;T_0})^2 \rangle - \langle\hat{n}_{T;T_0}\rangle}{\langle\hat{n}_{T;T_0}\rangle},
\end{align}
where $\hat{n}_{T;T_0} := \int_{T_0}^{T_0+T}ds\hat{b}^\dagger(s)\hat{b}(s)$ is the number operator for the beam over the interval $(T_0,T_0+T]$. This quantity may be interpreted as the normalised variance in the number of detections made by an ideal photodetector monitoring the beam over this time interval. For example, $-1\leq Q_{T;T_0} < 0$ would imply that the variance in photodetections made between successive measurements is less than the mean, which is the defining property of sub-Poissonian light.

For time-stationary fields, this quantity may be expressed in terms of the second-order Glauber coherence function~\cite{Davidovich,ZouMandel}
\begin{align}\label{Q-g2}
    Q_{T} = \frac{\langle\hat{n}_T\rangle}{T^2}\int_{-T}^{T}ds\left(T - |s|\right)\left( g_{\rm ps}^{(2)}(s) - 1 \right),
\end{align}
where $g_{\rm ps}^{(2)}(s):= g^{(2)}(T_0,T_0+s,T_0+s,T_0)$. Here, the subscript ps stands for photon statistics and we have dropped the redundant initial time $T_0$ from the notation for the case of a stationary field. Expressing $Q_T$ in this manner will be useful for calculations made in the following sections. Of particular interest to us is the long-time limit, where $T\xrightarrow{}\infty$, as this is the counting duration that gives the smallest $Q$-parameter. We therefore drop the subscript for further ease of notation, such that $Q := Q_{T\xrightarrow{}\infty}$ and use this quantity as our measure of the degree of sub-Poissonianity in the laser beam. 

\subsection{Conditions on the Laser and its Beam}\label{conditions}

In order to speak of the Heisenberg limit for the coherence of a CW laser beam, we must define the problem by specifying the constraints placed on the device and the beam it produces. In Ref.~\cite{HL}, the constraints which led to the derivation of the Heisenberg-limit were expressed as four precise conditions. To arrive at a more robust upper bound on $\mathfrak{C}$ that encompasses a more general class of laser models, we adopt the first three of these outright, while relaxing the fourth as much as possible such that an upper bound on the scaling law for $\mathfrak{C}$ with $\mu$ can still be derived based on the same methodology of proof. 

Together, these conditions encapsulate the most fundamental aspects of the radiation that is produced from the laser models developed throughout the 1960's and 70's~\cite{Louisell, SSL, wiseman2016}, while also being as general as possible about the mechanism by which this coherent beam is produced with the proviso that the laser does \textit{produce} the beam. That is, it cannot, for example, be an empty box through which a beam, from some other source, is shone (how this restriction is achieved will become clear). Instead, the conditions require one to place a box around all the devices and processes that are used to produce the coherent (and possibly sub-Poissonian) beam, and treat that as the laser device. Accordingly, our interchangeable use of the terms ``laser" and ``laser device" throughout this work is shorthand for ``the entire set of systems inside the box that carry coherent excitations whose phase give rise to the laser phase". 

Explicitly, the four conditions are as follows:

\begin{enumerate}
    \item \textbf{One Dimensional Beam}---The beam propagates away from the laser in one direction at a constant speed, occupying a single transverse mode and polarisation. The beam can therefore be described by a scalar quantum field with the annihilation operator $\hat{b}(t)$ satisfying $[\hat{b}(t),\hat{b}^\dagger(t')] = \delta(t-t')$.
\end{enumerate}

The notion of such a field characterized by the annihilation operator $\hat{b}(t)$ was used in the above definition of $\mathfrak{C}$. Essentially, $\hat{b}(t)$ represents the bit of the beam which was emitted from the device at time $t$, which propagates away from the laser at the speed of light.

\begin{enumerate}
    \addtocounter{enumi}{1}
    \item \textbf{Endogenous Phase}---Coherence in the beam proceeds from coherence in the excitations within the laser. Formally, a phase shift imposed on the
    laser state at some time $T_0$ will lead, in the future, to the
    same phase shift on the beam emitted after time $T_0$, as well
    as on the laser state.
\end{enumerate}

Imposing a phase shift at time $T_0$ can be described by the action of the superoperator $\mathcal{U}_{\rm c}^\theta:=\hat{U}_{\rm c}^\theta\bullet\hat{U}_{\rm c}^{\theta\dagger}$ on the laser state, with the unitary $\hat{U}^\theta_{\rm c}:=\exp{i\theta\hat{n}_{\rm c}}$ and the generator $\hat{n}_{\rm c}$ being the number operator for the excitations stored within the laser. While this operator must have non-negative integer eigenvalues, there is notably no assumption that the laser device consists of a single mode. Condition 2 requires that the effect of this phase shift at any later time $T_0+T$ can be described by the superoperator $\mathcal{U}_{\rm cb}^\theta=\hat{U}_{\rm cb}^\theta\bullet\hat{U}_{\rm cb}^{\theta\dagger}$ acting on the laser state as well as the beam segment generated over the interval $(T_0,T_0+T]$, where $\hat{U}^\theta_{\rm cb}:=\exp{i\theta(\hat{n}_{\rm c}+\hat{n}_{T;T_0})}$.

When considering bounds on the coherence, this condition is imperative as it guarantees that no external sources add phase information to to beam; if they did, they would be required to be treated as part of the laser and hence contribute to the mean excitation number, $\mu$, stored within the device. This number is guaranteed to exist by the third condition:

\begin{enumerate}
    \addtocounter{enumi}{2}
    \item \textbf{Stationary Beam Statistics}---The statistics of the laser and beam have
    a long-time limit that is unique and invariant under time
    translation.  This means that the mean excitation number within the laser, $\langle \hat{n}_{\rm c} \rangle$, has a unique stationary value $\mu$. 
\end{enumerate}

The fourth condition that we impose on the beam is similar to the ``Ideal Glauber$^{(1),(2)}$-coherence" condition of Ref.~\cite{HL}. This original condition stated that the first- and second-order Glauber coherence functions of Eq.~(\ref{n_corrs_norm}) for the laser beam be well-approximated by that of a coherent state undergoing pure phase diffusion (Eq.~(\ref{las_ideal})). By defining it in this manner, an upper bound for $\mathfrak{C}$ with a specific prefactor for the scaling law was able to be derived, namely $\mathfrak{C}\lesssim2.975\mu^4$. However, this condition is rather restrictive on the beam Glauber coherence properties; we therefore seek to relax this condition as much as possible, while still being able to obtain a bound on the scaling law for $\mathfrak{C}$. The fourth condition that we impose on the beam which achieves this is as follows:

\begin{enumerate}
    \addtocounter{enumi}{3}
    \item \textbf{Passably Ideal Glauber$^{(1),(2)}$ Coherence---}The first- and second-order Glauber coherence functions are \textit{passably close} to that produced by the state of an ideal laser, i.e, an eigenstate $\ket{\beta(t)}$ of $\hat{b}(t)$ with eigenvalue $\beta(t) = \sqrt{\mathcal{N}}e^{i\sqrt{\ell}W(t)}$ and $W(t)$ representing a Wiener process.
\end{enumerate}
Here, the addition of ``passably" in the above definition is to distinguish this condition from the original Condition 4. Explicitly, what we mean by our new Condition 4 is
\begin{subequations}
    \begin{align}\label{c4.1}
        |g_{\rm laser}^{(1)}(s,t)-g^{(1)}_{\rm ideal}(s,t)|= O(1),
    \end{align}
    \begin{align}\label{c4.2}
        |g_{\rm laser}^{(2)}(s,s',t',t)-g^{(2)}_{\rm ideal}(s,s',t',t)|= O(\mathfrak{C}^{-1/2}),
    \end{align}
\end{subequations}
for all values of the time arguments such that the difference between any two times is $O(\sqrt{\mathfrak{C}}/\mathcal{N})$. Consideration of this time difference is to yield the tightest upper bound for the coherence of such a laser (see Theorem 1, below). The subscripts ``ideal" and ``laser" seen in Eqs.~(\ref{c4.1}) and (\ref{c4.2}) denote the coherence functions pertaining to an ideal laser beam (i.e, that which is described exactly by Eq.~(\ref{las_ideal})), and those for a specific laser model, respectively.

The difference between the current Condition 4 and the original one conceived in Ref.~\cite{HL} is that the latter would replace $O$ on the RHS of Eqs.~(\ref{c4.1})~and~(\ref{c4.2}) with $o$. With Bachmann–Landau notation, $f(y) = o(g(y))$ formally means that for all positive constants, $k$, there exists a $y_0$, such that, for all $y>y_0$, then $|f(y)|<kg(y)$, while $f(y) = O(g(y))$ instead means $|f(y)|\leq kg(y)$, with $k$ and $y$ specified in the same manner~\cite{Burgisser}. The former condition is thus a much stricter requirement on the beam. The most important implication of this change is that this updated condition permits $g^{(2)}_{\rm laser}(s,s',t',t)$ to deviate considerably from an ideal laser beam, therefore including models with sub-Poissonian beam photon statistics. As we will see, this allows beams with corresponding values of $Q$ arbitrarily close to the minimum of $-1$ to satisfy these four conditions.

In order for Condition 4 to be meaningful, such that one may compare the coherence of given laser beam and that of an ideal beam, we define the ``linewidth", $\ell$, as
\begin{align}\label{define_ell}
    \ell := 4\mathcal{N}/\mathfrak{C},
\end{align}
with $\mathfrak{C}$ and $\mathcal{N}$ respectively given by Eqs.~(\ref{coh_general})~and~(\ref{flux}). This does not necessarily imply that this quantity $\ell$ corresponds to the FWHM of a Lorentzian power spectrum as suggested from the argument leading to Eq.~(\ref{coh_intuitive}). However, for the families of laser models that we consider in this paper, $g^{(1)}_{\rm laser}(s,t)$ will be shown to have an exponential decay to a good approximation, even in scenarios with maximal sub-Poissonianity in the beam. Therefore, we can still identify the quantity $1/\ell$ with the time in which it takes the phase inside the device to become fairly randomized. Thus the intuition we provide above for the coherence, that $\mathfrak{C}$ is the number of mutually coherent photons emitted into the beam, is still valid.

\section{The Heisenberg Limit for $\mathfrak{C}$}
\label{section3}

We now move on to present a proof of the upper bound for $\mathfrak{C}$, under the revised conditions discussed previously. This proof closely follows that for Theorem 1 in Ref.~\cite{HL}, with the difference here being that it is modified according to Eqs.~(\ref{c4.1})~and~(\ref{c4.2}), such that a more general Heisenberg limit for $\mathfrak{C}$ is derived. Therefore, this limit encompasses laser models that exhibit highly sub-Poissonian beam statistics. A consequence of this generalisation is that an upper bound with a specific prefactor is no longer able to be obtained, only the scaling $\mathfrak{C}=O(\mu^4)$, which is nevertheless sufficient to talk of the Heisenberg limit. The proof of the following theorem bears a good deal of resemblance with that of Ref.~\cite{HL} and we will therefore utilize the lemmas involved within that proof; these are stated explicitly in Appendix~\ref{lemmas}.

\textbf{Theorem 1:} (Generalisation of the upper bound on $\mathfrak{C}$). \textit{For a laser which satisfies conditions 1--4 stated above, the coherence is bounded from above:}
\begin{align}
    \mathfrak{C} = O(\mu^4),
\end{align}
\textit{with $\mu$, the mean number of excitations within the laser.}

\textit{Proof.} This involves an observer, Effie, to first perform a \textit{filtering} heterodyne measurement on the beam over the time interval $[T-\tau,T)$. A second observer, Rod, is then tasked with estimating the optical phase, $\phi_F$, that is encoded on the laser state at time $T$ by Effie's measurement. The methods that Rod considers to carry this out is to either make a \textit{retrofiltering} heterodyne measurement on the beam over the time interval $(T,T+\tau]$ to obtain the estimate $\phi_R$, or perform a direct measurement on the laser to obtain the estimate $\phi_D$. The proof works by verifying that the result $\phi_R$ cannot outperform $\phi_D$ as an estimate of $\phi_F$, such that an upper bound on $\mathfrak{C}$ will follow from known results on optimal covariant phase estimation~\cite{Bandilla1991}.

In order to describe the two heterodyne measurements in this problem, the unitary operators $e^{i\hat{\phi}_R}$ and $e^{i\hat{\phi}_F}$ are defined, which have unit-modulus complex eigenvalues and arguments that give the phase estimates $\phi_F$ and $\phi_R$ for the respective filtering and retrofiltering measurements. These operators may be expressed as~\cite{wiseman_milburn_2009}
\begin{subequations}\label{hetero1}
    \begin{align}
        e^{i\hat{\phi}_F} = \hat{F}/\sqrt{\hat{F}^\dagger\hat{F}},
    \end{align}
    \begin{align}
        e^{i\hat{\phi}_R} = \hat{R}/\sqrt{\hat{R}^\dagger\hat{R}},
    \end{align}
\end{subequations}
with
\begin{subequations}\label{hetero2}
    \begin{align}
        \hat{F} := \int_{T-\tau}^Tdtu_F(t)\hat{b}(t) + \hat{a}^\dagger_F,
    \end{align}
    \begin{align}
        \hat{R} := \int_T^{T+\tau}dtu_R(t)\hat{b}(t) + \hat{a}^\dagger_R.
    \end{align}
\end{subequations}
Here, $\hat{a}_{F(R)}$ are annihilation operators for the ancillary vacuum modes that enters into heterodyne detection and $u_{F(R)}$ are normalised filter functions, which, for simplicity, are taken as $u_F(t) = u_R(-t) = \tau^{-1/2}[H(t)-H(t-\tau)]$, where $H(t)$ is the Heaviside step function. By Lemma 4, we know that the operators $\hat{F}$ and $\hat{R}$ are \textit{phase covariant} (see Appendix~\ref{lemmas}).

First, with Condition 3 (Stationary Beam Statistics), we let the laser device be in its unique steady state, $\rho_{\rm c}^{\rm ss}$. If Condition 2 (Endogenous Phase) holds, then Lemma 2 implies that this steady state will be invariant under all optical phase shifts. Effie performs her filtering measurement to encode an optical phase, $\phi_F$, on the laser device corresponding to her measurement outcome. Because this measurement is phase covariant, and is being performed on a phase-invariant laser state, Lemma 1 applies. This means that the conditioned state of the laser following Effie's measurement is equivalent to a fiducial state (\ie, a state independent of the outcome $\phi_F$), $\rho_0$, with $\phi_F$ encoded by the generator $\hat{n}_c$. Crucially, this conditioned state, $\rho_{c|\phi_F} = e^{i\phi_F\hat{n}_c}\rho_0e^{-i\phi_F\hat{n}_c}$, has a mean excitation number $\mu$ regardless of the measurement outcome $\phi_F$; this is verified from Lemma 3.

We now consider Rod's heterodyne measurement. To compare how correlated this retrofiltering measurement result, $\phi_R$, is with Effie's result, $\phi_F$, we consider the quantity $1 - |\langle e^{i(\hat{\phi}_R - \hat{\phi}_F)} \rangle|^2$. This provides a convenient measure of the phase spread by recognizing that, for small errors $\theta$, $1-|\langle e^{i\theta} \rangle|\approx\langle\theta^2\rangle - \langle\theta\rangle^2$ is the mean-square error (MSE) for an unbiased estimate of $\theta$. With the definitions provided in Eqs.~(\ref{hetero1})~and~(\ref{hetero2}), it is possible to express this quantity as being asymptotically equivalent to an expression involving the first- and second-order Glauber coherence functions defined in Eqs.~(\ref{n_corrs}a)~and~(\ref{n_corrs}b)~\cite{HL}. That is, for $\ell\tau\ll1$ and $\mathcal{N}\tau\gg1$, one has
\begin{widetext}
    \begin{align}\label{eq long}
        \begin{split}
            1 - |\langle e^{i(\hat{\phi}_R - \hat{\phi}_F)} \rangle|^2 \sim & \frac{1}{2\mathcal{N}^2\tau^2} + \frac{1}{\mathcal{N}\tau^3}\int_0^\tau dt\int_0^\tau ds g^{(1)}(s,t) + \frac{1}{2\tau^4}\bigg{[} \int_0^\tau ds \int_{-\tau}^0 ds' \int_0^\tau dt' \int_{-\tau}^0 dt g^{(2)}(s,s',t',t) \\ & - \int_0^\tau ds \int_0^\tau ds' \int_{-\tau}^0 dt' \int_{-\tau}^0 dt g^{(2)}(s,s',t',t) \bigg{]}.
        \end{split}
    \end{align}
\end{widetext}
It is straightforward to evaluate Eq.~(\ref{eq long}) for an ideal laser described by the state $\ket{\beta(t)}$, as defined in Eq.~(\ref{las_ideal}). To leading order in $\ell/\mathcal{N}$, we find
\begin{align}\label{MSE id}
    1 - |\langle e^{i(\hat{\phi}_R - \hat{\phi}_F)}\rangle|^2_\textrm{ideal} \sim 2\sqrt{\frac{2\ell}{3\mathcal{N}}}.
\end{align} 
To arrive at this expression we have chosen $\tau = \sqrt{3\mathfrak{C}/2}/\mathcal{N}$, as this choice for the time interval minimizes the error in the retrofiltering measurement. This is done such that the tightest possible bound for the coherence is obtained from this method of proof. It is worth noting that while this prefactor of $\sqrt{3/2}$ has been included for completeness, it is irrelevant for what we require here. In fact, one could just as well set $\tau=\Theta(\sqrt{\mathfrak{C}}/\mathcal{N})$ and arrive at the same result. This is because, unlike Theorem~1 of Ref.~\cite{HL}, we are concerned only with the scaling of the upper bound for $\mathfrak{C}$, rather than obtaining a specific prefactor.

If one now considers an arbitrary laser model that satisfies Conditions 1--4, it may be seen from Equations~(\ref{eq long})~and~(\ref{MSE id}) that the \textit{relative} difference, $\Delta$, in the MSE, between this model and an ideal laser for the retrofiltering measurement, can be bounded above:
\begin{align}\label{rel diff}
    \begin{split}
        \Delta = & ~O\left(\max_{s,t\in[0,\tau]}|\delta g^{(1)}(s,t)|\right) \\ & + O\left(\sqrt{\mathcal{N}/\ell}\max_{s,s',t',t\in[-\tau,\tau]}|\delta g^{(2)}(s,s',t',t)|\right),
    \end{split}
\end{align}
where 
\begin{align}\label{delta_gn}
    \delta g^{(n)}(s_1,... , s_{2n}) := g^{(n)}_{\textrm{laser}}(s_1,... , s_{2n}) - g^{(n)}_{\textrm{ideal}}(s_1,... , s_{2n}).
\end{align}
From here we may write down an expression for an upper bound on the MSE from a retrofiltering measurement for such laser models,
\begin{align}\label{hetero ub}
    1 - |\langle e^{i(\hat{\phi}_R - \hat{\phi}_F)}\rangle|^2_\textrm{laser}\leq2\sqrt{\frac{2\ell}{3\mathcal{N}}}\left(1+|\Delta|\right).
\end{align}
Invoking Condition 4, the relative difference $\Delta$ becomes $O(1)$, and the RHS of Eq.~(\ref{hetero ub}) becomes $O(\sqrt{\ell/\mathcal{N}})$. Rewriting this in terms of the coherence, from Eq.~(\ref{define_ell}), we have
\begin{align}\label{bound1}
    1 - |\langle e^{i(\hat{\phi}_R - \hat{\phi}_F)}\rangle|^2_\textrm{laser}=O(\mathfrak{C}^{-1/2}).
\end{align}

Rod's second method of estimating $\phi_F$ is now considered. This involves performing a direct optical phase measurement on the laser. As stated above, the mean excitation number, $\mu$, of the laser is preserved after Effie performs filtering. This means that Lemma 5 can be applied, and from this we can identify from Eq~(\ref{lemma5}) $\hat{\phi}$ with $\hat{\phi}_D$ and $\bar{\phi}$ with $\phi_F$ plus the average phase value of the fiducial state $\rho_0$. Hence, we may write
\begin{align}\label{bound3}
    1 - |\langle e^{i(\hat{\phi}_D - \phi_F)} \rangle|^2 \gtrsim 4|z_A/3|^3\mu^{-2},
\end{align}
where $z_A\approx-2.338$ is the first zero of the Airy function~\cite{Bandilla1991}.

The direct measurement that would achieve equality in the above equation for the MSE would outperform any other measurement of the phase at time $T$. This is because the phase information imprinted onto the laser by Effie's filtering measurement is encoded only by the generator $\hat{n}_{\rm c}$, as shown by Lemma 1. We can therefore state that $\phi_D$ outperforms $\phi_R$ as an estimate of $\phi_F$. In terms of our measures of phase spread, this amounts to the mathematical statement
\begin{align}\label{bound2}
    1 - |\langle e^{i(\hat{\phi}_R - \hat{\phi}_F)} \rangle|^2 \geq 1 - |\langle e^{i(\hat{\phi}_D - \hat{\phi}_F)} \rangle|^2,
\end{align}
for $\mu\gg1$. The upper bound on $\mathfrak{C}$ then follows readily upon applying Eqs.~(\ref{bound1})~and~(\ref{bound3}) to Eq.~(\ref{bound2}), that is,
\begin{align}
    \mathfrak{C} = O(\mu^4).
\end{align}

\hspace*{\fill} $\qed$

It will be demonstrated numerically that $\mathfrak{C}\propto\mu^4$ for three specific families of laser models considered in this Paper. Asserting that this scaling is indeed at the Heisenberg limit, Eqs.~(\ref{c4.1})~and~(\ref{c4.2}) are required to be verified; this is also to be shown numerically.

\section{Numerical Methods}
\label{section4}

Calculating the observable quantities of interest, namely the coherence and $Q$-parameter, for the laser models that we are to introduce, requires the evaluation of correlation functions of the beam. Given the large size of the Hilbert space of these systems, evaluating such functions using standard numerical methods is a computationally expensive task. We use infinite Matrix product state (iMPS) techniques that have been developed over recent decades, which make simulating large systems more tractable by providing efficient ways of describing the entanglement content of the wavefunction (see, \eg, Refs.~\cite{MPS_rev1, MPS_rev2} for reviews). In this section we briefly review the treatment of a laser in the context of an MPS sequential quantum factory~\cite{schon1, schon2}, which is outlined in detail in Ref.~\cite{HL}, and discuss how particular physical quantities are calculated within this framework.

\subsection{iMPS of a Laser Beam}

\begin{figure}[H]
\includegraphics[width=1.0\columnwidth]{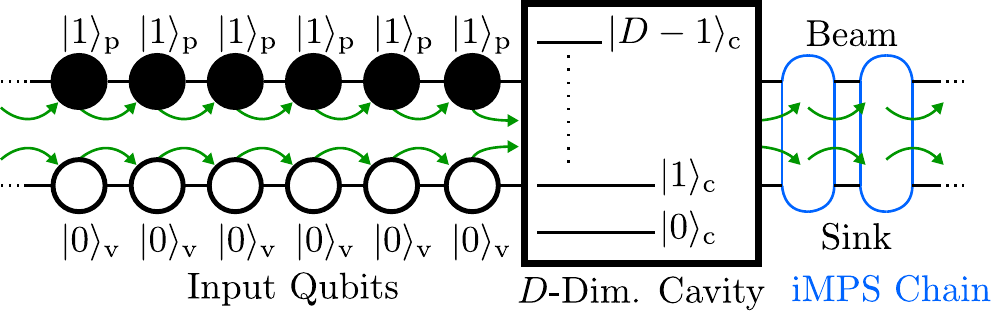}
\caption{\label{schematic1} Basic schematic of the laser model applicable to the $p$- and $p,\lambda$-families. A $D$-level cavity converts a pair of input qubits (pump and vacuum) into a pair of output qubits (beam and sink) at each-time step of duration $\delta t$. Green arrows indicate the movement of the input and output qubits from one time-step to the next. The chain of output qubits, of indefinite length, is described by an iMPS with bond dimension $D$, equal to the Hilbert space dimension of the laser cavity.}
\end{figure}

In order to describe a laser beam in an iMPS framework, it is necessary to discretize the laser process by which incoherent excitations pumped into a cavity are converted into coherent excitations within the output beam. To this end, we consider the most basic form of a laser system as shown in Fig.~\ref{schematic1}, which consists of five elements that are all essential for operation: a ``cavity" (c), pump (p), vacuum input (v), beam (b) and sink (s). In this model of a laser, the pump and vacuum inputs may be considered as a stream of incoming qubits into the cavity, which itself is treated as a $D$-level system with the non-degenerate number operator $\hat{n}_{\rm c} = \sum_{n = 0}^{D-1} = n|n\rangle_{\rm c}\langle n|$. The beam and sink are taken as a joint 4-level system (o), such that the laser consists of a single output in alignment with the requirements of an MPS sequential generation scheme. Addressing the beam alone is thus achieved by tracing over the sink. In this discretized approximation, a single beam qubit corresponds to an arbitrarily short beam segment of duration $\delta t$, such that it is occupied by at most one photon, where the bosonic operator for the beam is transformed as $\sqrt{\delta t} \hat{b} \xrightarrow[]{}\sigma^-_{\rm b}$, with $\sigma^-_{\rm b} = |1\rangle_{\rm b}\langle 0|$.

This discretized time evolution of the cavity and its outputs is governed by the generative interaction
\begin{align}
    \hat{V}_q = \sum_{j_{q+1},m,n}A_{mn}^{[j_{q+1}]}|m\rangle_{\rm c}\langle n|\otimes\ket{j_{q+1}}_{\rm o},
\end{align}
where $\ket{j_{q+1}}_{\rm o}:=\ket{\lfloor j/2 \rfloor_{q+1}}_{\rm b}\otimes\ket{(j\ \rm{mod}\ 2)}_{\rm s}$ is defined on the output space. This generative interaction corresponds to an isometry (a purity-preserving, completely-positive, trace-preserving map) from a $D$-dimensional vector space to a $4\times D$-dimensional one. The isometry condition, $\hat{V}^\dagger\hat{V} = I_{D}$, with $I_m$ as the $m\times m$ identity matrix, translates to a completeness orthonormality relation
\begin{align}\label{ortho_cond}
    \sum_{j=0}^3A^{[j]\dagger}A^{[j]}=I_{D}.
\end{align}

Of particular interest to us is the \textit{one-site unit-cell} infinite MPS (iMPS) that $\hat{V}$ eventually creates. In terms of the $A$ matrices, this is given by
\begin{align}
    \begin{split}
            & \ket{\Psi_{\rm MPS}} = \sum_{...,j_{q_0},j_{q_0-1},j_{q_0-2},...}\bra{\Phi(q=+\infty)}_{\rm c}...A^{[j_{q_0}]}_{(q_0)} \\ & A^{[j_{q_0-1}]}_{(q_0-1)}A^{[j_{q_0-2}]}_{(q_0-2)}...\ket{\Phi(q=-\infty)}_{\rm c}\ket{...,j_{q_0},j_{q_0-1},j_{q_0-2},...}.
    \end{split}
\end{align}
Here, $\ket{\Phi(q)}$ represents the state of the cavity at the discrete time $q$ and it is assumed that in the last step $q=+\infty$ the cavity decouples from the output. The $(q_i)$-subscripts shown above may also be dropped, which is permitted given that the outputs are translationally invariant. It is the case for all of our laser models that the largest-magnitude eigenvalue of the iMPS identity transfer matrix in its \textit{flattened space}, $\mathcal{T} = \sum_{j=0}^3A^{[j]*}\otimes A^{[j]}$, is non-degenerate. This ensures that a unique steady state of the cavity exists and additionally renders the boundary states, $\ket{\Phi(q=\pm\infty)}_{\rm c}$, to be irrelevant, in the sense that they will not appear in any calculations of the correlation functions. 

It is possible to relate this picture to one which is more familiar to a traditional quantum optics framework by defining a generative unitary interaction, $\hat{U}_{\rm int}$, according to 
\begin{align}
    \hat{V}\ket{\psi}_{\rm c} \equiv \hat{U}_{\rm int}(\ket{\psi}_{\rm c}\ket{1}_{\rm p}\ket{0}_{\rm v}).
\end{align}
Here, the unitary would act on the Hilbert space of the cavity and two input (pump and vacuum) qubits, to produce the evolved cavity state along with the two output (beam and sink) qubits. In doing so, this unitary describes the processes by which excitations are gained and subsequently lost from the laser system in a single, discretized time step of evolution. In the typical case, for standard laser systems, $\hat{U}_{\rm int}$ would describe a linear (in $\hat{a}$ and $\hat{a}^\dagger$) interaction between a harmonic oscillator storing the exciations ($[\hat{a},\hat{a}^\dagger]=1$) and its surrounding environment~\cite{wiseman1997}. The key change that is necessary to surpass the SQL for laser coherence is to make the interactions between the device and its environment highly nonlinear.

For the models considered in this paper, we impose conservation of energy on the unitary $\hat{U}_{\rm int}$, i.e, $\hat{n}_{\rm c} + \hat{n}_{\rm v} + \hat{n}_{\rm p} = \hat{n}'_{\rm c} + \hat{n}'_{\rm v} + \hat{n}'_{\rm p}$ (with primes denoting the operators following the application of $\hat{U}_{\rm int}$). In doing so, we ensure that Condition 2 is satisfied, which requires that all phase information imprinted on the beam proceeds only from the laser~\cite{HL}. This results in the $D\times D$ dimension $A$-matrices being highly sparse, where each matrix has at most a single non-zero diagonal, with $O(D)$ free parameters. For each matrix, these non-zero elements correspond to the entries $A^{[0]}_{m+1,m}$, $A^{[1]}_{m,m}$, $A^{[2]}_{m,m}$ and $A^{[3]}_{m,m+1}$, and are taken to be real and non-negative, consistent with a spectral peak at $\omega = 0$.

These $A$-matrices have clear physical interpretations: $A^{[0]}$ describes the gain process into the cavity, relating to the amplitude of the cavity receiving an uncorrelated photon from the pump without emitting a photon to the output. $A^{[1]}$ and $A^{[2]}$ relate to the amplitude of the process where the cavity receives a pump photon and sends it directly into the sink and beam, respectively. $A^{[2]}$ is therefore set to zero, as the photons emitted into the beam by this process add noise instead of contributing to the beam's coherence, therefore $A^{[1]} = \sqrt{I_D - A^{[0]\dagger}A^{[0]} - A^{[3]\dagger}A^{[3]}}$ in accordance with Eq.~(\ref{ortho_cond}). Finally, $A^{[3]}$ describes the process of laser loss, which creates the beam, and relates to the amplitude of the cavity receiving an uncorrelated pump photon and emitting a single photon to both the beam and the sink, de-exciting the cavity by a single level.

It is furthermore possible to draw a connection between the laser dynamics in the framework presented above and that of a master equation for the cavity state, $\dot{\rho} = \mathcal{L}\rho$. This may be seen by considering the cavity state evolved by a single time step $\delta t$ in the iMPS framework, 
\begin{align}\label{iMPS evolution}
    \rho(t+\delta t) = \sum_j A^{[j]}\rho(t)A^{[j]\dagger}.
\end{align}
By taking the length of this discrete time interval to be infinitesimal, $\delta t\xrightarrow{}0^+$, a master equation is obtained with the Liouvillian, $\mathcal{L}$, taking the form
\begin{align}\label{liouvillian}
    \begin{split}
        \frac{d\rho}{dt} & = \mathcal{L}\rho \\ &
         = \mathcal{N}\left(\mathcal{D}[\hat{G}] + \mathcal{D}[\hat{L}]\right)\rho,
    \end{split}
\end{align}
where $\mathcal{D}[\hat{c}] := \hat{c}\bullet\hat{c}^\dagger - \frac{1}{2}(\hat{c}^\dagger\hat{c}\bullet + \bullet\hat{c}^\dagger\hat{c})$ is the usual Lindblad superoperator, and $\hat{G}$ and $\hat{L}$ are ``gain" and ``loss" operators, respectively, which specify how energy is added and released from the cavity. These gain and loss operators directly correspond to the iMPS operators $B^{[0]}:=A^{[0]}/\sqrt{\gamma}$ and $B^{[3]}:=A^{[3]}/\sqrt{\gamma}$, respectively, with $\gamma = \mathcal{N}\delta t$. Explicit forms of these are given in Section~\ref{Families}.

\subsection{iMPS Calculations of $\mathfrak{C}$, $Q$ and Glauber$^{(1),(2)}$ Correlators}

For the discretized laser model introduced above, calculating the coherence amounts to evaluating the quantity
\begin{align}\label{coh_discrete}
    \mathfrak{C} = \sum_{q' = -\infty}^{\infty}\langle \sigma^+_{\rm b}(q+q')\sigma^-_{\rm b}(q) \rangle.
\end{align}
The method that is employed here to achieve this, as well as the calculation of the first- and second-order Glauber correlation functions, is the well-established method of manipulating MPS transfer operators~\cite{MPS_rev1}. In Ref.~\cite{HL} it was demonstrated that, from this method, Eq.~(\ref{coh_discrete}) may be re-expressed as
\begin{align}\label{cohmps}
    \mathfrak{C} = -2(1|(B^{[3]*}\otimes I_D)\cdot{\rm inv}(\mathbb{Q}\mathbb{L}\mathbb{Q})\cdot (I_D\otimes B^{[3]})|1),
\end{align}
which is written in \textit{flattened space}, where superoperators such as the transfer-type operators become $D^2\times D^2$-sized matrices and $D\times D$-sized operators are transformed into flattened $D^2$-sized vectors. Under this notation, $\rm{inv}(\bullet)$ represents the matrix inverse operation, $(1|\leftrightarrow{}I_D$ and $|1)\leftrightarrow{}\rho_{\rm ss}$ are the left- and right-leading eigenvectors of $\mathcal{T}$, both of which have their eigenvalues equal to unity. This implies that the latter eigenvector satisfies the steady-state equation $\sum_j\hat{A}^{[j]}\rho_{\rm ss}\hat{A}^{[j]\dagger} = \rho_{\rm ss}$. Finally, $\mathbb{Q}=I_{D^2} - |1)(1|$, and $\mathbb{L}$ represents the flattened space version of the superoperator defined in Eq.~(\ref{liouvillian}).

A simplified expression for the Q-parameter may also be found in terms of this flattened space iMPS language. Starting from Eq.~(\ref{Q-g2}), $Q$ may be re-expressed as
\begin{align}\label{Qmps}
    \begin{split}
        Q = 2\gamma(1| & (B^{[3]*}\otimes B^{[3]}) \\ & \cdot{\rm inv}(I_{D^2}-\mathbb{Q}\mathcal{T}\mathbb{Q})\cdot(B^{[3]*}\otimes B^{[3]})|1).
    \end{split}
\end{align}

Additionally, the first- and second-order Glauber coherence functions are
\begin{subequations}
    \begin{align}\label{G1mps}
        G^{(1)}(s,0) = (1|(B^{[3]*}\otimes I_D)\mathbb{E}(s)(I_D \otimes B^{[3]})|1),
    \end{align}
    \begin{align}\label{G2mps}
        \begin{split}
            G^{(2)}(s,s',t',t) = (1|(B^{[3]*}\otimes I_D)\mathbb{E}(s'-s)(B^{[3]*}\otimes I_D) \\ \mathbb{E}(t'-s')(I_D \otimes B^{[3]})\mathbb{E}(t-t')(I_D \otimes B^{[3]})|1),
        \end{split}
    \end{align}
\end{subequations}
where $\mathbb{E}(t):=\exp\{\mathcal{N}t\mathbb{L}\}$. It is also worth noting that Eq.~(\ref{G2mps}) is for the specific time ordering $s<s'<t'<t$, but other time orderings may be calculated in a similar manner following an appropriate permutation of the bosonic operators. Eqs.~(\ref{G1mps})~and~(\ref{G2mps}) will be of use when verifying that our laser models satisfy Condition 4, details of which are given in Appendix~\ref{verification}.

\section{Comments on Practicality}\label{V}

As we have touched upon in the previous section, in order for a laser model to produce a beam with Heisenberg-limited scaling of $\mathfrak{C}$, the interaction between the cavity and its environment must be highly nonlinear. In theory, this can be achieved by choosing atypical forms of operators $\hat{G}$ and $\hat{L}$ entering into the Lindblad terms of Eq.~(\ref{liouvillian})~\cite{HL}. While the structure of these are similar to the standard creation and annihilation operators, in the sense that they will raise and lower the excitation number by one, respectively, their coefficients in the number basis of the cavity would be considerably different from the usual $\bra{n-1}\hat{a}\ket{n}=\sqrt{n}$.

Engineering the unitary $\hat{U}_{\rm int}$ that would give rise to these dynamics has been show to be achievable in principle on a circuit-QED architecture~\cite{HL}. In that work, this was demonstrated by considering two qubits that mediate an interaction between a microwave resonator and its environment. Specifically, by having a standard Jaynes-Cummings type interaction between both qubits and the resonator, the required exotic gain and loss processes may be controlled by modulating the detuning of these qubits from cavity resonance. Although this model is relatively straightforward to describe mathematically, it would be technically challenging to implement on near-term experimental hardware and was developed primarily as a proof-of-principle to demonstrate achievability of the Heisenberg limit for $\mathfrak{C}$. A potential alternative route to implement these dynamics, which may be more feasible, would be to work with a similar circuit-QED architecture, but instead exploiting the disperse shift between a series of microwave resonators and artificial atoms, and employing reservoir engineering techniques similar to those utilized in Ref.~\cite{Gertler_2021}.

As we have shown in Ref.~\cite{Ostrowski} and will also proceed to demonstrate in this paper, there is no reason that these engineered systems should produce a beam that exhibits Poissonian beam photon statistics. This is unlike regular laser devices, where the Poissonian photon statistics of the beam is largely inherently ensured due to the standard gain mechanisms and linearity in $\hat{a}$ of cavity loss. To yield such beam statistics would require exact tailoring of the gain and loss processes that are specified by $\hat{G}$ and $\hat{L}$, respectively. To attest to this, we refer the reader to the red curve of Fig.~2(c) in Ref.~\cite{Ostrowski} for an example. This plots the behaviour of the Mandel-$Q$ parameter for a particular family of Heisenberg-limited laser models as a function a model parameter that modifies the flatness of the coefficients (in the number basis) of the gain and loss operators. There, we see that there is there is nothing special about the Poissonain case ($Q=0$) within that curve. In particular, it is neither a maximum or a minimum, so a deviation in the operators that give rise to that particular value of $Q$ will lead to a deviation in the photon statistics of the same size, and deviations in the sub-Poissonian direction seem just as likely as deviations in the super-Poissonian direction. 

The current proofs of the upper bound for $\mathfrak{C}$ in a given device require constraints to be placed on the photon statistics. In any practical system, the controls that would give rise to the desired engineered unitary $\hat{U}_{\rm int}$ are of course expected to be susceptible to technical imperfections. These imperfections would cause deviations from the desired gain and loss processes and in turn, given the reasoning presented in above, deviations away from the Poissonain beam photon statistics. This means that when considering ultimate achievable bounds on the production of coherence by a laser device, it is important, both from theoretical and practical perspectives, to account for the photon statistics of a given laser model in order to assert that it meets the criterion for the beam to be considered Heisenberg-limited. 


\section{Families of Laser Models}\label{Families}

\subsection{Quasi-Isometric, Markovian Gain ($p$-family)}

In this section, we introduce three families of laser models, each of which will be shown to to produce a beam with Heisenberg-limited scaling of $\mathfrak{C}$ for some range of parameter values. The first family may be characterized by the master equation
\begin{align}\label{mastereqp}
    \begin{split}
        \frac{d\rho}{dt} & = \mathcal{L}_{\rm M}^{(p,0)}\rho \\ & = \mathcal{N}\left( \mathcal{D}[\hat{G}^{(p,0)}] + \mathcal{D}[\hat{L}^{(p,0)}] \right)\rho,
    \end{split}
\end{align}
describing the evolution of a $D$-level cavity introduced in the preceding section. Here, the subscript ``M" in the Liouvillain superoperator stands for ``Markov", $p\in(0,\infty)$, and the non-zero elements of the gain and loss operators are defined generally as
\begin{subequations}\label{GainLoss}
    \begin{align}\label{gain_gen}
        G^{(p,x)}_n \propto \left(\frac{\sin{\left(\pi\frac{n+1}{D+1}\right)}}{{\sin{\left(\pi\frac{n}{D+1}\right)}}}\right)^\frac{px}{2},
    \end{align}
    \begin{align}\label{loss_gen}
        L^{(p,x)}_n \propto \left(\frac{\sin\left(\pi\frac{n}{D+1}\right)}{{\sin\left(\pi\frac{n+1}{D+1}\right)}}\right)^{\frac{p(1-x)}{2}},
    \end{align}
\end{subequations}
which are expressed here in the number basis of the cavity for $(0<n<D)$, where
\begin{align}
    G^{(p,x)}_{n} \equiv \bra{n}\hat{G}^{(p,x)}\ket{n-1}, \quad L^{(p,x)}_{n} \equiv \bra{n-1}\hat{L}^{(p,x)}\ket{n}.
\end{align}
The parameter $x$ in Eqs.~(\ref{GainLoss}a--b) can be any real number; however, for the case at hand we set $x=0$ therefore imposing a ``flat" gain operator with $\hat{G}^{(p,0)}_n\propto1$.

For this Master Equation~(\ref{mastereqp}), the steady state cavity photon distribution may be found with $\rho_n = |G^{(p,x)}_n/L^{(p,x)}_n|^2\rho_{n-1}$, such that
\begin{align}\label{ss_dist}
    \rho_n = \alpha \sin^p{\left(\pi\frac{n+1}{D+1}\right)}, \quad (0\leq n<D),
\end{align}
where $\rho_n = \bra{n}\rho_{\rm ss}\ket{n}$ and $\mathcal{L}_{\rm M}^{(p,0)}\rho_{\rm ss}=0$. For this distribution, the mean photon number is $\mu=(D-1)/2$ and in the asymptotic limit, as $D\xrightarrow{}\infty$, the normalization factor has a straightforward expression,
\begin{align}
    \lim_{D\xrightarrow{}\infty}D\alpha = \sqrt{\pi}\frac{\Gamma\left(\frac{2+p}{2}\right)}{\Gamma\left(\frac{1+p}{2}\right)}.
\end{align}

This family of laser models, defined by Eqs.~(\ref{mastereqp})~and~(\ref{GainLoss}), is a generalisation of the original laser model that exhibited Heisenberg-Limited coherence. That model was found via an optimization of the iMPS $A$-matrices to maximize $\mathfrak{C}$, for $D$ finite, which suggested the Ansatz~(\ref{ss_dist}) with $p=4$~\cite{HL}. This value of $p$ was also used in the companion Letter~\cite{Ostrowski} for the additional two families of models discussed below. In introducing this additional parameter, $p$, we are now able to modify the variance of the steady-state cavity distribution and thus explore the implications of this on the various physical properties of the system. Comparing this with the two other families of models that are to be introduced, we may distinguish this one by taking note of the gain process into the laser. In this situation, we identify this family as having a \textit{quasi-isometric, Markovian} gain. Recalling that an isometric operator, $\hat{V}$, requires $\hat{V}^\dagger\hat{V}=\hat{I}_D$, we deem the use of the term \textit{``quasi-isometric"} to be appropriate because $\hat{G}^{(p,0)\dagger}\hat{G}^{(p,0)} = \hat{I}_D - \hat{\Pi}_{\rm top}$, where $\hat{\Pi}_{\rm top}$ is the projector onto the upper-most cavity state,
\begin{align}
    \hat{\Pi}_{\rm top}:=\ketbra{D-1}{D-1}.
\end{align}
Therefore, in taking the limit where the dimension of the Hilbert space of the cavity tends to infinity, $\hat{G}^{(p,0)}$ is approximately isometric for the particular types of states we concern ourselves with in this work. That is, for states with negligible population in the upper-most cavity state in this large-$D$ limit. This point is addressed in more detail in Appendix~\ref{RP_Appendix}. Throughout this Paper, this family of laser models will be referred to as the \textit{p-family}.

\subsection{Non-Isometric, Markovian Gain ($p,\lambda$-family)}

The second family of laser models we introduce may also be characterized by a master equation in the form of Eq.~(\ref{liouvillian}), with the Liouvillian $\mathcal{L}_{\rm M}^{(p,\lambda)}$, such that
\begin{align}\label{mastereqlam}
    \begin{split}
        \frac{d\rho}{dt} & = \mathcal{L}_{\rm M}^{(p,\lambda)}\rho \\ & = \mathcal{N}\left( \mathcal{D}[\hat{G}^{(p,\lambda)}] + \mathcal{D}[\hat{L}^{(p,\lambda)}] \right)\rho.
    \end{split}
\end{align}

\begin{figure}[h]
\includegraphics[width=1.0\columnwidth]{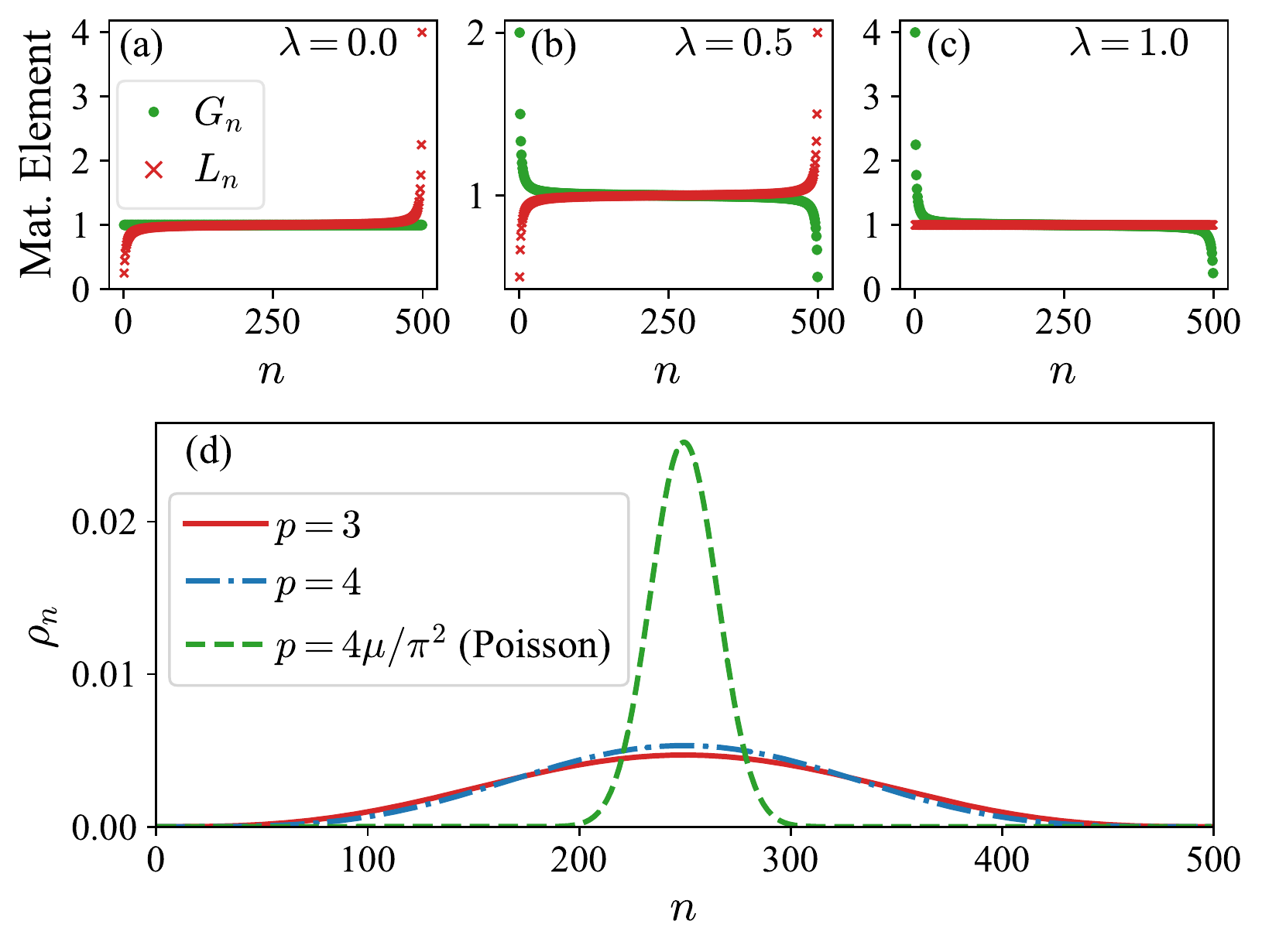}
\caption{\label{fig:2} (a-c) Non-zero matrix elements of the gain (green dots) and loss (red crosses) operators for the laser family exhibiting a non-isometric, Markovian gain model (\ie, the $p,\lambda$-family), for various choices of $\lambda$ and cavity dimension $D=500$. (d) Steady-state cavity photon distribution of the three laser families~(\ref{ss_dist}) for $p=3$ (solid red line), $p=4$ (blue dashed-dotted line) and  $p=4\mu/\pi^2$ (green dashed line), with cavity dimension $D=500$. Optimal coherence is obtained in each family of models for $p\approx4.15$.}
\end{figure}

In addition to $p$, we have introduced another continuous parameter, with $x\xrightarrow{}\lambda\in\mathbb{R}$, which allows for a significant modification to the form of the gain and loss operators of the system, while preserving the steady-state cavity distribution given in Eq.~(\ref{ss_dist}). Although $\lambda$ may take any value within the set of real numbers, of particular interest to us are values falling within the range $0\leq\lambda\leq1$. Here, one may interchange smoothly between a flat gain operator, with $\lambda\xrightarrow{}0.0$, to a flat loss, with $\lambda\xrightarrow{}1.0$. This relationship between $\lambda$ and the gain and loss operators is depicted in Fig.~\ref{fig:2}a--c, while the influence of $p$ on the cavity distribution may be seen in in Fig.~\ref{fig:2}d. As will be shown both numerically and from heuristic arguments, modification of the gain and loss operators in this manner will have notable implications on both the coherence and photon statistics of the system. This family of laser models, exhibiting in general a \textit{non-isometric} ($\hat{G}^{(p,\lambda)\dagger}\hat{G}^{(p,\lambda)}\not\approx \hat{I}_D$ for $\lambda\neq0$), \textit{Markovian} gain mechanism, will be referred to as the \textit{$p,\lambda$-family} according to the two key parameters which characterize it.

\subsection{Quasi-Isometric, Non-Markovian Gain ($p,q$-family)}

For the $p,\lambda$-family introduced above, we will demonstrate that the beam exhibits sub-Poissonian photon statistics under particular choices of parameter, but the maximum attainable reduction in photon noise in the output field (that is, $Q=-1$) is not achieved within this family. However, it is possible for complete noise reduction to be attained, at least in principle, for laser systems by introducing a mechanism by which the pumping of excitations into the cavity is done so in a regular manner~\cite{Golubev1984, Machida1987, Richardson1990,rp2,rp3,rp4,wiseman1993}. The basic idea behind this can be understood simply by a consideration of particle number conservation. That is, if every pumped excitation will sooner or later be emitted as a photon into the beam, then the long-time photon statistics of the output field will mirror that of the pumping mechanism. Hence, for a completely regular pump, complete photon noise reduction would be possible in the output field for long counting intervals~\cite{rp2,rp3,Yamamoto1992}.

Over the years a number of models have been proposed which can achieve this. For example, by making modifications to the Scully and Lamb theory~\cite{ScullyLamb} to account for a regular stream of excited atoms into the laser (see, \eg, Ref.~\cite{Walls_Milb} and references therein). Similar effects can be achieved through internal mechanisms by which pump electrons are recycled through a number of rate-matched energy levels~\cite{RalphSavage, RZGWalls}. To conclude this section we derive a master equation for a laser which incorporates a regular pumping mechanism, with the goal of producing a family of models that exhibit both Heisenberg-limited coherence, as well as complete photon noise reduction in the output field for long counting intervals for a specific choice of parameter. Our derivation follows closely to that given in Ref.~\cite{wiseman1993}; the aim here is to incorporate a continuous parameter into the model that allows for a smooth interchange between a Poissonian and regular injection of excitations into the $D$-level laser cavity mentioned above. 

\begin{figure}[H]
\includegraphics[width=1.0\columnwidth]{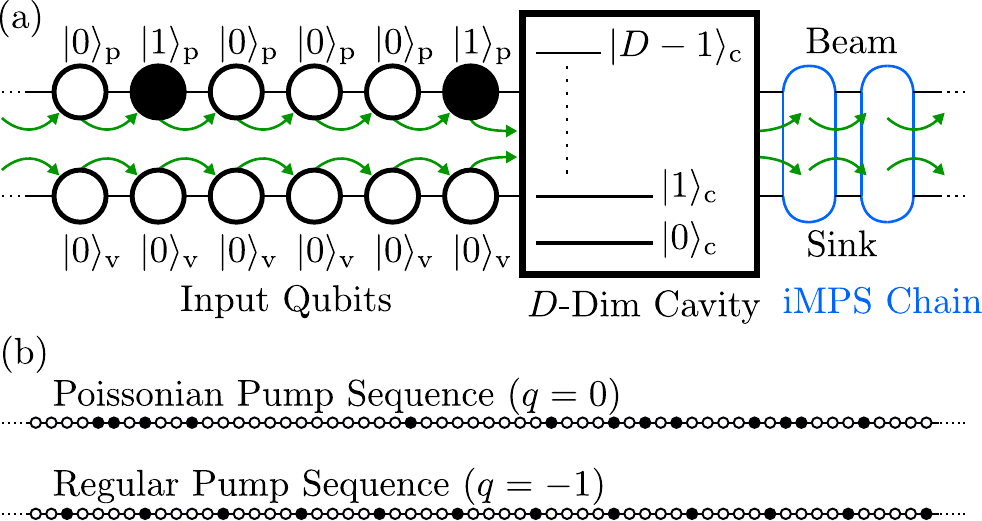}
\caption{\label{schematic2}(a) Basic schematic of the laser model applicable to the $p,q$-family. The elements of this model are much the same as that of Fig.~\ref{schematic1}. However, in this situation, the pumping sequence has been altered in order to facilitate the process by which pump excitations are deposited in the laser in a regular manner. (b) Qualitative depictions of particular examples of pumping sequences. Top row shows the pumping sequence for the choice of parameter $q=0.0$, where the number of excitations absorbed by the cavity in some time $\Delta t$ is sampled from a Poisson distribution. Bottom row shows the pumping sequence where the entry of excitations into the cavity are made to be perfectly regular by choosing $q=-1.0$. Here, there is no variance in the number of excitations absorbed by the cavity in the time $\Delta t$.}
\end{figure}

In order to aptly model this process within an iMPS framework, modifications must necessarily be made to the laser model introduced in the previous section. This modified system is depicted in Fig.~\ref{schematic2}a. Comparing this with Fig.~\ref{schematic1}, the most notable change is that that the pump qubits are no longer exclusively in the excited $\ket{1}_{\rm p}$ state. Although this is the case, this scenario represents a much more energy efficient process as it is formulated in such a way so that essentially every excited pump qubit will be converted into an excitation within the cavity. In particular, we define two unitary operators, $\hat{U}_{\rm loss}$ and $\hat{U}_{\rm gain}$, which evolve system by a single time step, $\delta t$, given that the pump qubit is in the state $\ket{0}_{\rm p}$ or $\ket{1}_{\rm p}$, respectively. Like the generative unitary interaction for the previous iMPS laser model, these two unitary operators map the $4\times D$-dimensional vector space consisting of the pump and vacuum input qubits, and the cavity, to a $4\times D$-dimensional one corresponding to the beam and sink generated by the cavity in $\delta t$, along with the evolved cavity state. 

Explicitly, these operators are
\begin{subequations}
\begin{align}
    \begin{split}
        \hat{U}_{\rm loss} = \exp \Big{\{} \sqrt{\gamma} \big{(} & \hat{L}^{(p,-q/2)}\otimes\hat{I}_{\rm p}\otimes\hat{\sigma}_{\rm v}^+ \\ & - \hat{L}^{\dagger(p,-q/2)}\otimes\hat{I}_{\rm p}\otimes\hat{\sigma}_{\rm v}^- \big{)} \Big{\}},
    \end{split}
\end{align}
\begin{align}
    \begin{split}
        \hat{U}_{\rm gain} = & \hat{G}^{(p,0)}\otimes\hat{\sigma}_{\rm p}^-\otimes\hat{I}_{\rm v} + \hat{G}^{\dagger(p,0)}\otimes\hat{\sigma}_{\rm p}^+\otimes\hat{I}_{\rm v} \\ & + \hat{\Pi}_{\rm top}\otimes|1\rangle_{\rm p}\langle 1|\otimes\hat{I}_{\rm v} + \hat{\Pi}_{\rm bot}\otimes|0\rangle_{\rm p}\langle 0|\otimes\hat{I}_{\rm v}.
    \end{split}
\end{align}
\end{subequations}
Here, $\hat{I}_{\rm p}$ and $\hat{I}_{\rm v}$ are $2\times2$ identity matrices acting on the space of the input pump and vacuum qubits, respectively. $\hat{\Pi}_{\rm bot}$ is the projector onto the ground state of the cavity, $\ket{0}_{\rm c}$. The non-zero elements of the gain and loss operators ($G^{(p,0)}_{n}$ and $L^{(p,-q/2)}_{n}$) are the same as those given in Eq.~(\ref{GainLoss}a--b), where we set $x\xrightarrow{}0$ and $x\xrightarrow{} -q/2\in(-\infty,1/2)$, respectively. Defining $G^{(p,0)}_{n}$ in this manner ensures that the process of gain into the cavity corresponds to a completely-positive, trace-preserving operation, while $L^{(p,-q/2)}_{n}$ is defined in this manner to preserve the steady state cavity distribution~(\ref{ss_dist}) (see Appendix~\ref{RP_Appendix} for details).

Considering generally a mixed cavity state, $\rho(t)$, the action of each of these gain and loss unitary operators lead to the incrementally evolved states
\begin{subequations}\label{gain_loss_superops}
\begin{align}
    \rho_{\rm loss}(t+\delta t) = (1 + \gamma\mathcal{D}[\hat{L}^{(p,-q/2)}])\rho(t),
\end{align}
\begin{align}
    \rho_{\rm gain}(t+\delta t) = (1 + (\mathcal{G}-1))\rho(t),
\end{align}
\end{subequations}
respectively, where $\mathcal{G}\bullet=\hat{G}^{(p,0)}\bullet\hat{G}^{\dagger(p,0)}+\hat{\Pi}_{\rm top}\bullet\hat{\Pi}_{\rm top}$. To arrive at these equations, we have traced over the beam and sink, as well as neglected terms of order $O(\gamma^{3/2})$ and higher. 

From here, we are in a position to incorporate pumping statistics which differ from the standard Poissonian case. To this end, we consider a short time, $\Delta t$, such that a large number, $n(\Delta t)$, of pump excitations have entered the cavity. Let us express this quantity in the following way
\begin{align}
    n(\Delta t) = \mathcal{N}\Delta t + \sqrt{\mathcal{N}(q+1)}\Delta W,
\end{align}
where $\Delta W$ represents a Wiener increment~\cite{wiseman1993} and $q$ is the Mandel-Q parameter of the gain process. That is, $q\xrightarrow{}0$ corresponds to a Poissonian process and $q\xrightarrow{}-1.0$ corresponds to a completely regular process (see Fig~\ref{schematic2}b for a depiction of each of these scenarios). The change to the cavity state as a result of the gain process in this time is then given by
\begin{align}\label{justgain}
    \rho_{\rm gain}(t+\Delta t) = \left(1 + \mathcal{D}[\hat{G}^{(p,0)}] + \mathcal{D}[\hat{\Pi}_{\rm top}]\right)^{n(\Delta t)}\rho(t).
\end{align}
Here, we are treating the gain and loss processes independently. This is justified under the assumption that $1\ll\mathcal{N}\Delta t\ll D$, as the action of the superoperators $\mathcal{D}[\hat{L}^{(p,-q/2)}]$ and $(\mathcal{G}-1)$ on the particular cavity states we consider in this work are of order $O(D^{-1})$. Details regarding this assumption may be found in Appendix~\ref{RP_Appendix}. Furthermore, given that the action of these superoperators is small, a binomial expansion of Eq.~(\ref{justgain}) is permitted, which, to second order in $(\mathcal{G}-1)$, is
\begin{align}\label{binom_exp}
    \begin{split}
            \rho_{\rm gain}(t+\Delta t) \approx & \big{[}1 + n(\Delta t)(\mathcal{G}-1) \\ & + (1/2)n(\Delta t)(n(\Delta t)-1)(\mathcal{G}-1)^2\big{]}\rho(t).
    \end{split}
\end{align}
Averaging over the uncertainty in the number of excitations added to the cavity and taking the limit $\Delta t \xrightarrow{} 0^+$, one obtains the master equation
\begin{align}\label{mastereqq}
    \begin{split}
        \frac{d\rho}{dt} & = \mathcal{L}_{\rm NM}^{(p,q)}\rho \\ & = \mathcal{N}\left(\mathcal{D}[\hat{G}^{(p,0)}] +\frac{q}{2}\mathcal{D}[\hat{G}^{(p,0)}]^2 + \mathcal{D}[\hat{L}^{(p,-q/2)}]\right)\rho,
    \end{split}
\end{align}
where we have also reinstated the loss term, $\mathcal{D}[\hat{L}^{(p,-q/2)}]$. Here, the subscript ``NM" stands for ``Non-Markovian". In order to maintain consistency with Eq.~(\ref{mastereqp}), such that Eq.~(\ref{mastereqq}) reduces to the master equation for the $p$-family for $q\xrightarrow{}0$, we have also let $(\mathcal{G} - 1) \approx \mathcal{D}[\hat{G}^{(p,0)}]$. Strictly speaking, one should have $(\mathcal{G} - 1) = \mathcal{D}[\hat{G}^{(p,0)}]+\mathcal{D}[\hat{\Pi}_{\rm top}]$, however, though this neglected term is significant in general, it may be ignored by considering the same argument as before. That is, because the edge elements for the class of states considered in this study are negligible, the contribution of $\mathcal{D}[\hat{\Pi}_{\rm top}]$ in Eq.~(\ref{mastereqq}) will also be negligible. The reader is again referred to Appendix~\ref{RP_Appendix} for details.

Equation~(\ref{mastereqq}) is only an approximate description of a regularly pumped laser, being a Markovian equation describing a generally non-Markovian process. However, equations of the same form as this have commonly been employed in laser theory~\cite{Golubev1984,rp3,wiseman1993,Walls_Milb,Bergou_Phase} and we will indeed show that reasonable results are obtained from our subsequent analysis. This family of laser models can be thought of exhibiting a \textit{quasi-isometric, non-Markovian} gain mechanism. In line with our notation for the other two families of laser models, we will refer to this as the \textit{$p,q$-family}.

\section{Coherence and Sub-Poissonianity}
\label{Numerical_Analysis}

In this section, we employ Eqs.~(\ref{cohmps}) and~(\ref{Qmps}) to compute the quantities $\mathfrak{C}$ and $Q$, respectively, for our three families of laser models~(\ref{mastereqp}),~(\ref{mastereqlam})~and~(\ref{mastereqq}). In doing so, we systematically explore the parameter space that is characteristic to each family to ascertain the influence that these parameters have on the coherence and Mandel-$Q$ parameter for each of the laser beams. We note that throughout this section, and for the remainder of this Paper, we let $\mathcal{N} = 1$ to set a convenient unit for time.

\begin{figure}[H]
\includegraphics[width=1.0\columnwidth]{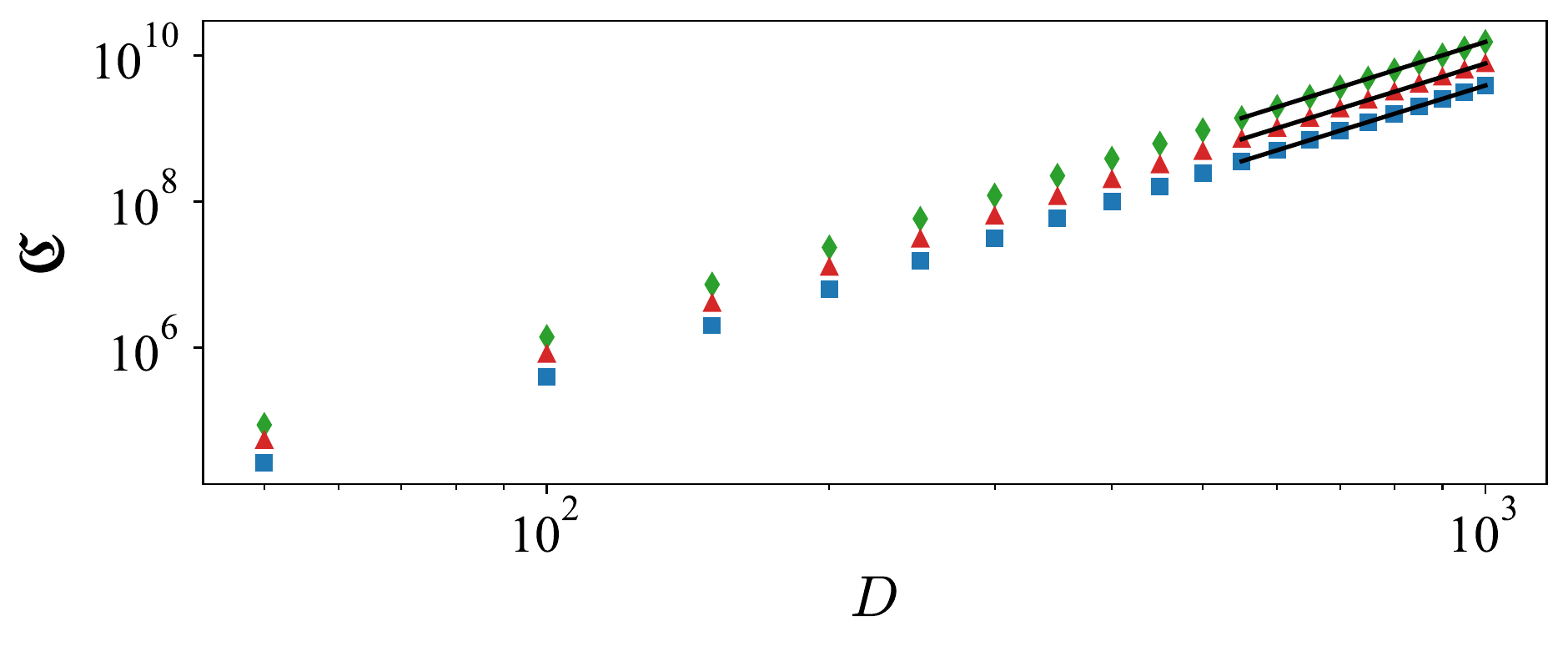}
\caption{\label{fig:C_D} iMPS calculations of the beam coherence for our three families of laser models as a function of dimension $D=2\mu+1$. Squares correspond to the laser family with the partially-isometric, Markovian gain model ($p$-family); triangles to the non-isometric, Markovian gain model ($p,\lambda$-family); and diamonds to the partially-isometric, non-Markovian gain model ($p,q$-family). Parameters are chosen such that beam coherence is maximized, these are $p = 4.1479$, $(p,\lambda) = (4.1479,0.5)$ and $(p,q) = (4.1479,-1.0)$ for each family, respectively. Solid black lines are fits to the data for $D\in[550,1000]$, assuming a power law. Respectively, these are $\mathfrak{C} = 0.0040D^4$, $\mathfrak{C} = 0.0082D^4$ and $\mathfrak{C} = 0.0140D^4$.}
\end{figure}

As an initial result, we show that Heisenberg-limited scaling for the coherence may be realized in each of our three families of laser models. Fig.~\ref{fig:C_D} shows iMPS calculations of $\mathfrak{C}$ for each family plotted against the ``cavity" dimension, $D$, ranging between $D=50$ and $D=1000$. In this plot, the particular choice of the other parameters are $p=4.15$ for the $p$-family, $p=4.15$ and $\lambda=0.5$ for the $p,\lambda$-family, and $p=4.15$ and $q=-1.0$ for the $p,q$-family. These parameter values are those which maximize $\mathfrak{C}$ in each family. 
Fitting a power law $\mathfrak{C}=c\mu^w$ (recalling that $\mu = (D-1)/2$) indicates that the coherence scales with the fourth power of $\mu$ in each case. By construction, each of these families of laser models satisfy the first three conditions on the laser and its beam (that is, One Dimensional Beam, Endogenous Phase and Stationary Statistics)~\cite{HL}. In Appendix~\ref{verification}, we verify numerically that Eqs.~(\ref{c4.1}) and~(\ref{c4.2}) are satisfied, therefore demonstrating that Condition 4 is also fulfilled. From this, we may say that for these particular parameter values, these laser families all perform at the Heisenberg limit. Moreover, we find that the largest coherence is attained within the $p,q$-family, with the prefactor $c$ in this optimized model being approximately four times larger than that for the $p$-family, and approximately twice as large as that for the $p,\lambda$-family.

\begin{figure}[t]
\includegraphics[width=1.0\columnwidth]{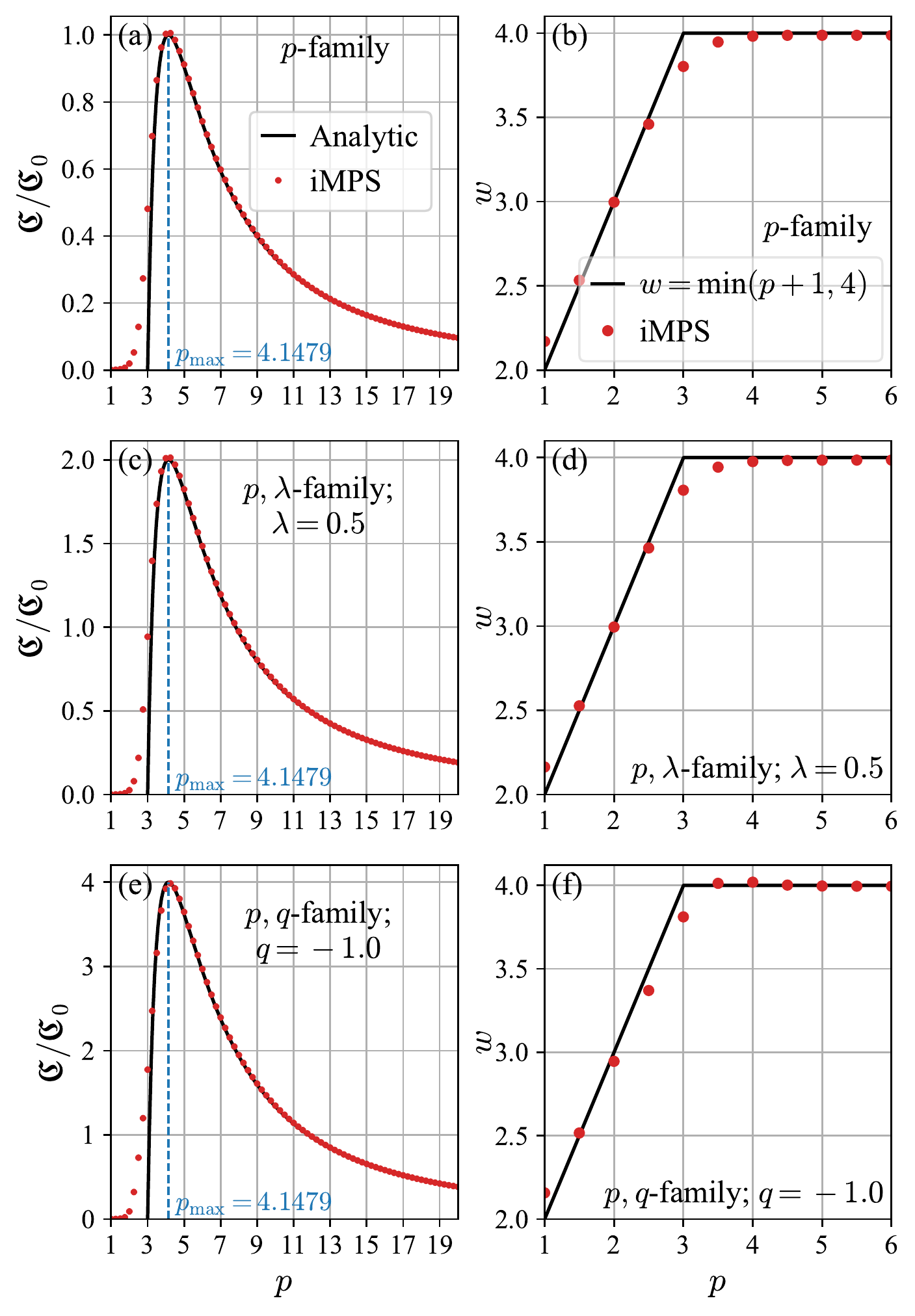}
\caption{\label{fig:5} (a,c,e): iMPS calculations of the beam coherence (red dots) for the $p$-, $p,\lambda$- and $p,q$-families of laser models described in the text as a function of $p$ for a cavity dimension $D=1000$. These values for the coherence have been normalized to the maximum value of that obtained from the $p$-family, for which $p=4.1479$. Solid black curves show the corresponding formulae for each family of models given by Eqs.~(\ref{analytic_c0}), (\ref{coh_lam_general}) and (\ref{coh_q_general}). Additional choice of parameters for the sub-Poissonian laser families are $\lambda = 0.5$ for the $p,\lambda$-family and $q=-1.0$ for the $p,q$-family, which yield maximum photon noise reduction in the beam. (b,d,f): Exponent $w$ (red dots) from the result of fitting $\mathfrak{C}=c\mu^w$ for particular choices of $p$, obtained for iMPS calculations up to bond dimension $D=1000$ for the three families of laser models shown in the left-hand panels of this figure. Solid black lines depict $w = \min(p+1,4)$ as a guide for the eye.}
\end{figure}

Looking at influence of the characteristic parameters in more detail now, we first direct our focus towards the $p$-family, which is the most straightforward of the three. Bringing one's attention to Fig.~\ref{fig:5}a, a number of qualitative statements can be made with regard to the behaviour of the coherence for this family from this plot. The red dots in Fig.~\ref{fig:5}a show iMPS calculations of the coherence normalized to its maximal value, $\mathfrak{C}_0$, as a function of the parameter $p$ for a ``cavity" of fixed dimension $D=1000$. As $p$ is increased from its lowest value, a rapid increase in the coherence is seen before reaching its maximum value at $p\approx4.1479$. Increasing $p$ above this optimal value, the coherence decreases monotonically from $\mathfrak{C}_0$ at a relatively gradual rate. Strictly speaking, the location of the peak for the coherence has a small dependence on $D$, however from the analysis presented in Section \ref{general_p} we find that $p=4.1479$ maximizes the coherence in the asymptotic limit $D\xrightarrow{}\infty$. Aside from this, the overall qualitative behaviour of the coherence seen in Fig.~\ref{fig:5}a was found to be independent of $D$ for sufficiently large values of this parameter. Moreover, apart from a multiplicative factor, this behaviour with respect to $p$ is preserved in the $p,\lambda$- and $p,q$-families regardless of $\lambda$ and $q$ (see Figs.~\ref{fig:5}c~and~\ref{fig:5}e). Determining these multiplicative factors is the subject of the following section within this Paper, where analytical methods are employed to compute the coherence for each of these families of laser models.

This behaviour of the coherence with respect to relatively low values of $p$ seen in Figs.~\ref{fig:5}a,c,e can be elucidated by considering what is shown in Figs.~\ref{fig:5}b,d,f. There, each data point (red dots) indicate, for a given value of $p$, the exponent $w$ in a power law fitted to $\mathfrak{C}$ when evaluated as a function of $\mu$. For instance, the plots in Fig.~\ref{fig:C_D} show a particular example of this for $p=4.15$ and similar raw data is used to determine each data point shown in Figs.~\ref{fig:5}b,d,f for the various values of $p$ considered. We again make the point to emphasize that these plots are independent of the parameters $\lambda$ and $q$, and therefore the scaling of the coherence with $\mu$ for each family depends only on $p$. In particular, we find that two distinct regimes may be identified for the behaviour of the coherence with respect to $p$. For values $p\lesssim3$, the coherence appears to be sub-Heisenberg-limited, with the exponent roughly obeying the formula $w=p+1$. This explains the rapid change in the coherence that is seen in Figs.~\ref{fig:5}a,c,e for these relatively low values of $p$. On the other hand, for values $p\gtrsim3$, $w$ becomes constant and the coherence is Heisenberg-limited, being proportional to $\mu^4$. Although values $p>6$ are not shown here, it was found that this scaling is preserved for larger choices of $p$. This implies that the monotonic decrease in the coherence as a function of $p$ above the optimal value of $p\approx4.1479$ seen in Figs.~\ref{fig:5}a,c,e is a result of the specific form of prefactor $c$ in the power law $\mathfrak{C}=c\mu^w$, and not the exponent $w$. Heuristic arguments to explain the behaviour of $w$ between these two distinct regimes are provided in Section~\ref{general_p}.

\begin{figure}[t]
\includegraphics[width=1.0\columnwidth]{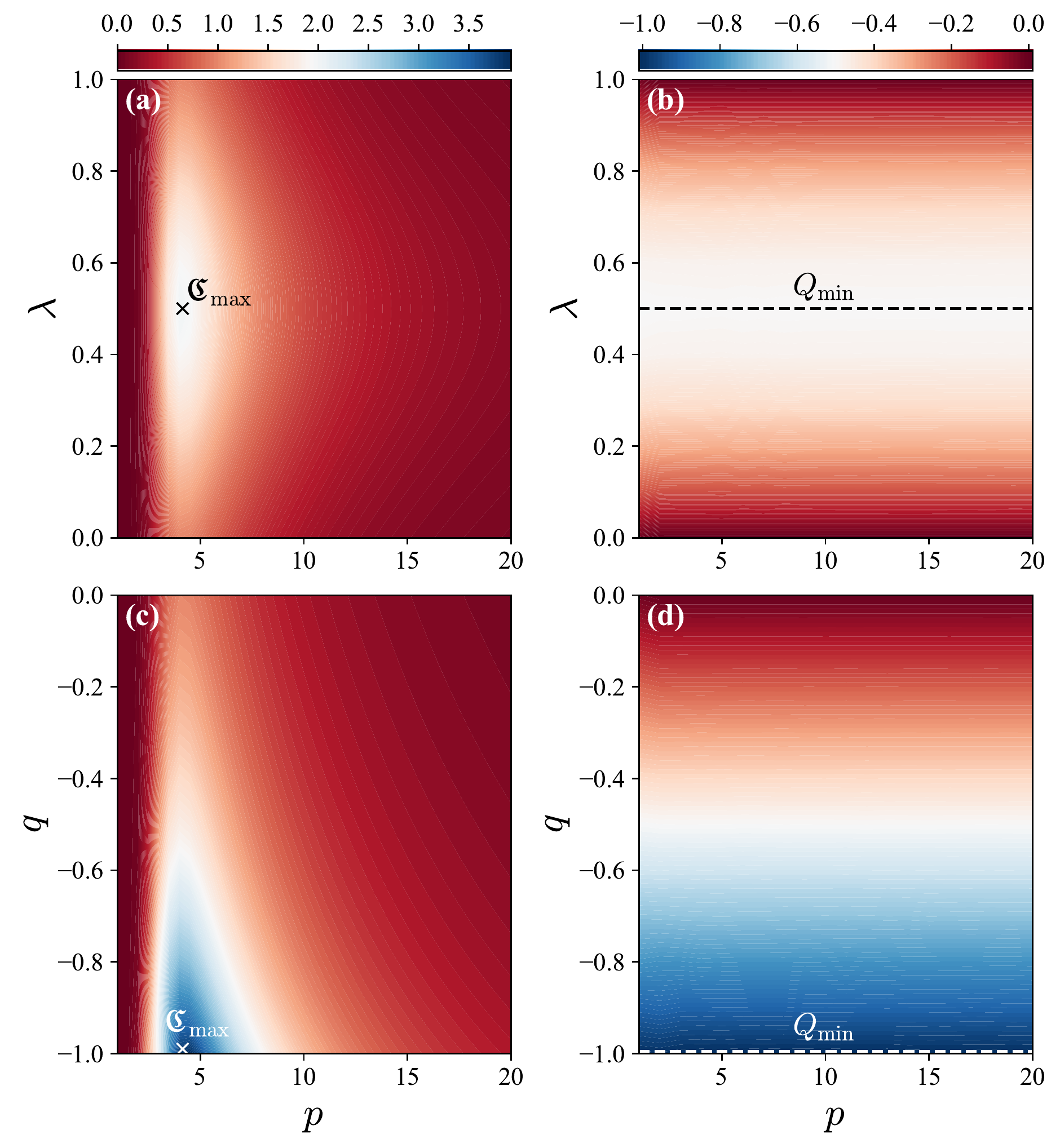}
\caption{\label{fig:3} (a,c): iMPS calculations of beam coherence for the $p,\lambda$-family and $p,q$-family, respectively, and with cavity dimension $D=1000$. These values of the coherence have been normalized to the maximum of that obtained for the $p$-family of lasers for $D=1000$. Crosses show the parameter choices which yield maximum coherence, according to Eqs.~(\ref{coh_lam_general}) and (\ref{coh_q_general}), which are located at $(p,\lambda) = (4.1479,0.5)$ and $(p,q) = (4.1479,-1.0)$, respectively. (b,d): Same as that shown for (a,c), but with normalized coherence substituted with the Mandel-$Q$ parameter of the beam. Dashed lines show the parameter choices which yield minimum $Q$ values, being $Q=-0.5$ and $Q = -1.0$ for the $p,\lambda$-family and $p,q$-family, respectively.}
\end{figure}

It is more interesting, however, to view the behaviour of the coherence in conjunction with that of the $Q$-parameter of the output field shown in Fig~\ref{fig:3}. In Fig~\ref{fig:3}a we provide a density plot of the coherence as a function of $p$ and $\lambda$ for the $p,\lambda$-family for $D=1000$. Recall that for $\lambda=0.0$, the $p,\lambda$-family reduces to the $p$-family. In this case, the system exhibits a flat gain such that $\hat{G}^{(p,\lambda=0)}$ is essentially a finite version of the Susskind-Glogower operator~\cite{Susskind}. Assuming a cavity state with negligible coefficient in its top level, $\ket{n=D-1}$, the action of this operator will preserve the phase statistics. Therefore, essentially no phase noise will be added via the gain process in this scenario~\cite{Wiseman1999}. On the other hand, Fig~\ref{fig:3}a shows that for a given value of $p$, the coherence is maximized by choosing $\lambda=0.5$, for which $\mathfrak{C}$ is twice that for the $p$-family with the same value of $p$. In this situation, a symmetry is imposed on the system where the matrix elements for the gain and loss operators are defined as reciprocals to one another. Given this observation, it is apparent that reducing the phase noise induced by the loss mechanism at the expense of that induced by the gain, to an extent, is advantageous to increase the coherence. 

In Fig.~\ref{fig:3}b, we plot $Q$ for the $p,\lambda$-family against the same parameters in Fig.~\ref{fig:3}a. This shows that the photon statistics are independent of $p$, and that the beam for the $p$-family is always characterized by Poissonian photon statistics ($Q=0.0$). We however find that $Q$ is minimized to a value of $-0.5$ when $\lambda=0.5$, which corresponds to a $50\%$ reduction in the number fluctuations of the laser beam below the shot noise limit for long counting intervals. This minimum value of $Q$ is found at exactly the same values which maximize the coherence within this family of laser models. Given this apparent ``win-win" situation between $\mathfrak{C}$ and $Q$ for the $p,\lambda$-family leads us to the conclusion that \textit{there is no trade-off between coherence and sub-Poissonianity for Heisenberg-limited lasers}.

As we have pointed out earlier, this minimum of $Q=-0.5$ seen for the $p,\lambda$-family can be surpassed and complete noise reduction for long times ($Q=-1.0$) is achievable by employing a regular pumping mechanism within the laser. This is exactly what we demonstrate in Fig.~\ref{fig:3}d, which displays the $Q$ parameter for our $p,q$-family against the parameters $p$ and $q$, again with $D=1000$. As anticipated, we find that the $Q$ parameter of the output field mirrors exactly that of $q$, the Mandel-$Q$ parameter of the pumping mechanism. For completely regular pumping, $q\xrightarrow{}-1$, we find $Q\xrightarrow{}-1$ (regardless of $p$), corresponding to $100\%$ reduction in the photon number fluctuations in the beam for long counting times. In addition to this, for a given $p$, we find that $\mathfrak{C}$ is maximized with $q\xrightarrow{}-1$, which is approximately four times that for the $p$-family for the same value of $p$ (see Fig.~\ref{fig:3}c). This takes the results for the $p,\lambda$-family a step further, by demonstrating that this synergistic effect between coherence and sub-Poissonianity in the output field persists in Heisenberg-limited lasers, even when extreme measures are taken to completely eliminate the photon noise in the beam. These results together mark a generalisation of the findings from previous studies relating to lasers with linear-loss \cite{wiseman1993,Ralph2004,Bergou_Phase,Benkert_Phase}, where we have extended the notion that a trade-off between $\mathfrak{C}$ and $Q$ does not necessarily exist, even for lasers with a phase diffusion rate that is far smaller than any laser model previously studied. Moreover, these results show that there is a ``win-win" relationship between these two quantities, as minimizing $Q$ comes hand-in-hand with a maximization of $\mathfrak{C}$.

Another observation to make here is with regard to the correlation time of the photon statistics. For the sub-Poissonian laser models we consider, the correlation time of $g^{(2)}_{\rm ps}(\tau)$ is much shorter than the coherence time, such that a significant degree of sub-Poissonianity in the beam may be observed for counting intervals where the phase of the beam remains fairly well-defined. This can be seen by considering both the $p,\lambda$- and $p,q$-families under the choice of parameters which minimize $Q$; respectively, these are $\lambda\xrightarrow{}0.5$ and $q\xrightarrow{}-1$ (regardless of the value of $p$). It is shown in Appendix~\ref{verification}, in both scenarios, that $1-g^{(2)}_{\rm ps}(0)=\Theta(\mathfrak{C}^{-1/2})$. Assuming an exponential decay for these functions (this is not strictly true for relatively small values of $p$, but is reasonable to assume for this qualitative argument), obtaining values of $Q=\Theta(1)$ with these parameter choices, as was demonstrated above, requires the correlation time to be $\Theta(\mathfrak{C}^{1/2})$ in accordance with Eq.~(\ref{Q-g2}). Interestingly, this is of the same order as optimal time for the filtering/retrofiltering measurements used in Theorem~1 (recalling the choice $\mathcal{N}=1$), which is a factor of $\mu^{-2}$ shorter than the coherence time of these laser models. This detail regarding the photon statistics of these laser families will be explored in more depth in future work.

\section{Generalized Formulae for Laser Coherence}
\label{general_p}

In this final section of results, we derive formulae for the laser coherence of our three families of laser models for a large range of values for $p$. These formulae are shown to be valid in the asymptotic limit $D\xrightarrow{}\infty$ for values $p\gtrapprox3$, for which Heisenberg-limited scaling, $\mathfrak{C}=\Theta(\mu^4)$, is achieved as indicated in Figs.~\ref{fig:5}b,d,f. In addition to this, although expressions for $\mathfrak{C}$ that are valid for $p\lessapprox3$ are unable to be obtained here, this analysis does give insight to the change in the exponent for the power law $\mathfrak{C}=c\mu^w$ to sub-Heisenberg-limited scaling for these parameter values.

\subsection{$p$-family}

\begin{figure}[H]
\includegraphics[width=1.0\columnwidth]{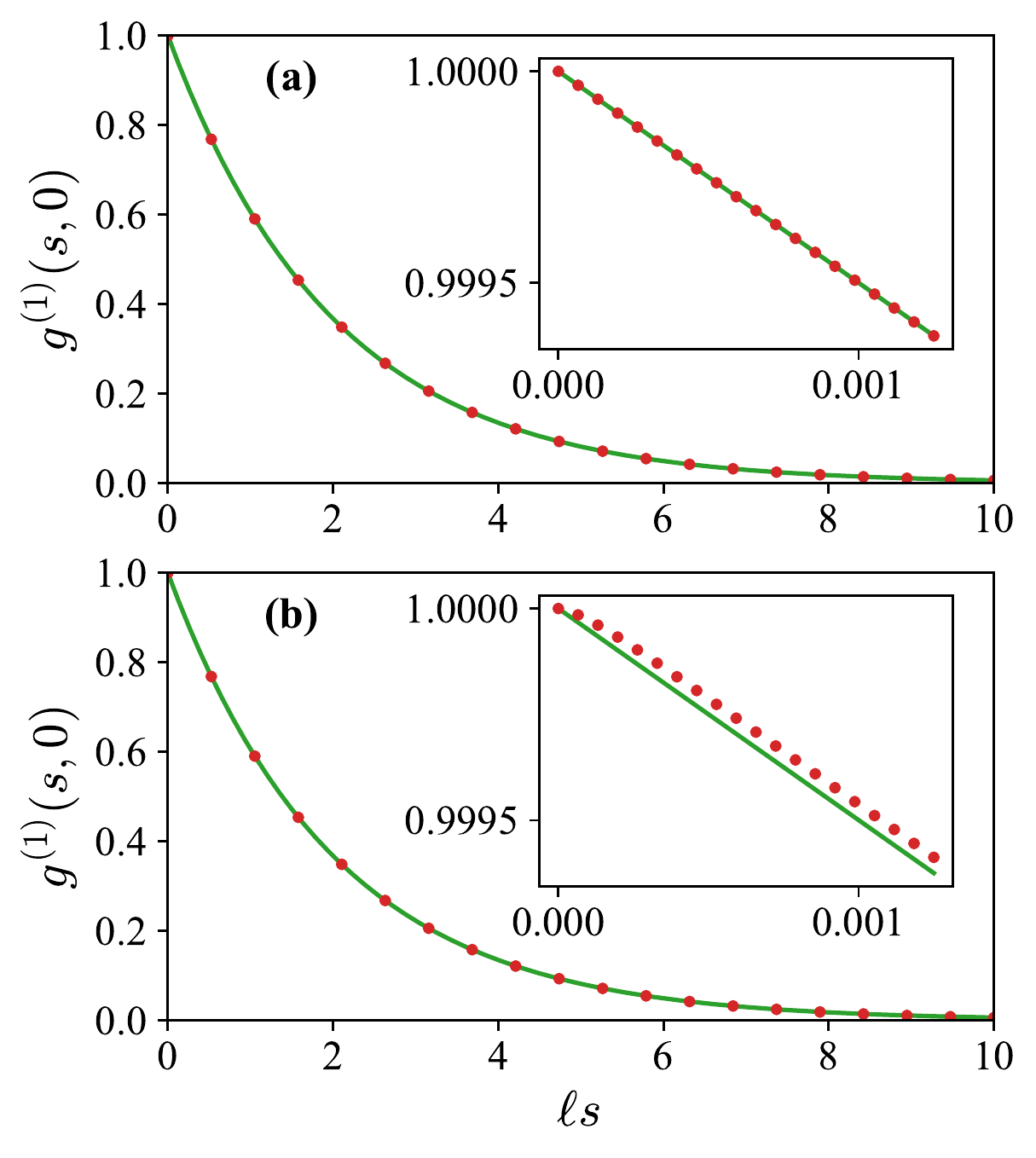}
\caption{\label{fig:8} (a): iMPS calculations (red dots) of the first-order Glauber coherence function $g^{(1)}(s,0)$ for the $p$-family of laser models, with $p=4.15$ and $D = 300$ over 10 coherence times, $1/\ell$. Green lines show the first-order Glauber coherence function for an ideal laser state, $g^{(1)}(s,0) = \exp(-\ell|s|/2)$. The inset give the same as what is shown in the larger panel, yet over a much shorter time scale such that $\exp(-\ell|s|/2)\approx-\ell|s|/2$. (b): Same as that shown in (a), but for the $p,\lambda$-family of laser models where $\lambda = 0.5$. We have omitted showing this for the $p,q$-family of laser models as these results are essentially the same as those shown for the $p,\lambda$-family in panel (b), where deviations occur between the laser model's and ideal laser's $g^{(1)}(s,0)$ behaviour for very short times when sub-Poissonian photon statistics are imposed on the beam.}
\end{figure}

We begin by focusing on the $p$-family, which exhibits Poissonian beam photon statistics. In order to derive an expression for the laser coherence, we appeal to the fact that the first-order Glauber coherence function is well-approximated by that of an `ideal' laser state given by Eq.~(\ref{las_ideal}), that is,
\begin{align}\label{g1_phase_diff}
    g^{(1)}(t) \approx \exp\left( -\ell|t|/2 \right).
\end{align}
Moreover, we assume that this holds to a first order approximation for an arbitrarily short time interval $\delta t$, which allows one to write
\begin{align}\label{g1_phase_diff_appx}
    \textrm{Tr}\left[ \hat{L}^{(p,0)\dagger} \mathcal{L}_{\rm M}^{(p,0)}\left(\hat{L}^{(p,0)}\rho_{ss}\right) \right] \approx -\frac{\ell}{2}.
\end{align}
This is justified on inspection of Fig.~\ref{fig:8}a, which shows numerical calculations of $g^{(1)}(t)$ overlapped with the function $e^{-\ell|t|/2}$, with $\ell = 4\mathcal{N}/\mathfrak{C}$ computed numerically. For short times, we see that the approximation of Eq.~(\ref{g1_phase_diff_appx}) is reasonable.

\begin{figure}
\includegraphics[width=1.0\columnwidth]{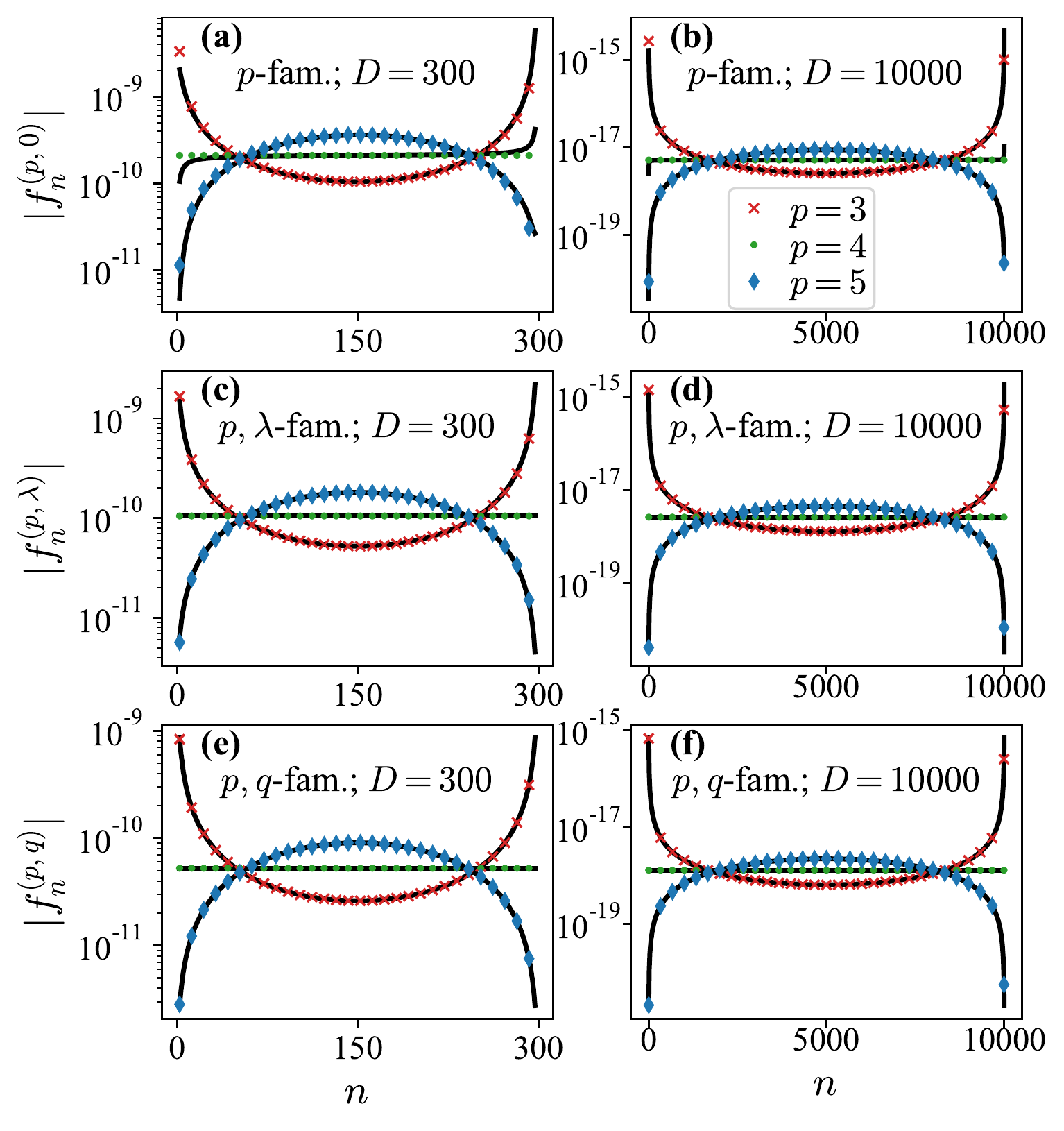}
\caption{\label{fig:7} (a): Absolute values of the diagonal elements defined by $f^{(p,0)}_n = [ \hat{L}^{(p,0)\dagger} \mathcal{L}_{\rm M}^{(p,0)}(\hat{L}^{(p,0)}\rho_{ss})]_{n,n}$ for cavity dimension $D=300$. Solid black lines show the exact elements, while markers indicate the elements obtained from the approximation given in Eq.~(\ref{approx_elem}). Respectively, red crosses, green dots and blue diamonds correspond to the cases of $p = 3,4,5$, respectively. Note that only a subset of equally spaced elements are marked here for clarity. (b): Same as for (a), but with an increased cavity dimension to $D=10000$. (c-d): Same as (a-b), respectively, but for the sum defined by $f_n^{(p,\lambda)} = (\mathcal{L}_{\rm M}^{(p,\lambda)}\varrho^{\phi=0}_c)_{n,n+1}$ and with $\lambda=0.5$. (e-f): Same as (a-b), respectively, but for the sum defined by $f_n^{(p,q)} = (\mathcal{L}_{\rm NM}^{(p,q)}\varrho^{\phi=0}_c)_{n,n+1}$ and with $q=-1$.}
\end{figure}

Using the photon-number basis $\{\ket{n}\}$ to evaluate the trace in Eq.~(\ref{g1_phase_diff_appx}), we can write it as the sum
\begin{align}\label{fn_define}
    \sum_{n=0}^{D-1}f^{(p,0)}_n \equiv \textrm{Tr}[ \hat{L}^{(p,0)\dagger} \mathcal{L}_{\rm M}^{(p,0)}(\hat{L}^{(p,0)}\rho_{ss})].
\end{align}
The elements $f_n^{(p,0)}$ may be written in terms of the elements of the steady-state cavity distribution,
\begin{align}\label{cases}
    f^{(p,0)}_n = 
    \begin{cases}
        0 & n = 0, \\
        \frac{-\rho_0^2}{2\rho_1} & n = 1, \\
        \frac{\rho_{D-2}}{2}\left(\sqrt{\frac{\rho_{D-3}}{\rho_{D-2}}} - \sqrt{\frac{\rho_{D-2}}{\rho_{D-1}}}\right)^2 - \frac{\rho_{D-2}}{2} & n = D-1, \\
        \frac{-\rho_{n-1}}{2}\left(\sqrt{\frac{\rho_{n-2}}{\rho_{n-1}}} - \sqrt{\frac{\rho_{n-1}}{\rho_{n}}}\right)^2 & {\rm otherwise}.
    \end{cases}
\end{align}
If one can evaluate the sum given by Eq.~(\ref{fn_define}), then an estimate for the linewidth of the laser can be obtained. This does not appear trivial at first glance. Luckily, significant simplifications can be made by investigating the behaviour of $f^{(p,0)}_n$ against the parameter $p$, which is depicted in Figs.~\ref{fig:7}a~and~\ref{fig:7}b for two different choices of cavity dimension. The solid black lines in these plots show the exact elements of the sum for values of $p = 3,4,5$. From this, along with Eq.~(\ref{cases}), we can identify two distinct regimes based on the dominant terms in the sum and how they change with $p$ for $D\gg1$. In particular, terms near the midpoint ($n\approx\mu$) are of order $O(D^{-5})$ and, roughly speaking, there are $O(D)$ elements that scale in this manner. Therefore, it would be expected that together these terms would contribute a quantity of order $O(D^{-4})$ to the total sum. On the other hand, we find that the edge elements have a greater dependence on the parameter $p$, where in the extreme cases (\eg, $n=1$ and $n= D-1$) they are of order $O(D^{-(p+1)})$---note the sudden ``flip" in the plots in moving from $p>4$ to $p<4$ as these edge elements suddenly become the largest terms in $f^{(p,0)}_n$. The first of these regimes is therefore identified for values $p>3$, where the sum is dominated by the elements around the midpoint, while the second is identified for values $p<3$, where the edge elements are instead dominant. As the coherence is directly proportional to the inverse of $\sum_nf^{(p,0)}_n$ (assuming that Eq.~(\ref{g1_phase_diff_appx}) holds), from this behaviour one should expect a change in the scaling of $\mathfrak{C}$ with $\mu$ when moving between these regimes. This competition between the central and edge elements of $f^{(p,0)}_n$ therefore explains the behaviour observed in Fig.~\ref{fig:5}b, as the exponent changes from $w\approx p+1$ for $p<3$, to $w\approx4$ for $p>3$.

Guided by this insight, we now attempt to evaluate Eq.~(\ref{fn_define}) for $p>3$ by considering the elements $1<n<D-1$, and performing a Taylor series expansion of $f^{(p,0)}_n$ about the point $\pi/(D+1) = 0$, which, to leading order in $\pi/(D+1)$, is
\begin{align}\label{approx_elem}
    \begin{split}
        f^{(p,0)}_n \approx & -\frac{\pi^{9/2}p^2}{8(D+1)^5}\frac{\Gamma\left(\frac{2+p}{2}\right)}{\Gamma\left(\frac{1+p}{2}\right)} \\ & \cdot\left(1+\cot^2\left(\pi\frac{n+1}{D+1}\right)\right)^2\sin^p\left(\pi\frac{n+1}{D+1}\right).
    \end{split}
\end{align}
The validity of this approximation can be visualised in Figs.~\ref{fig:7}a~and~\ref{fig:7}b, where the marked points correspond to an evenly-spaced subset of the elements within the approximation of Eq.~(\ref{approx_elem}). These are plotted on top of the exact elements of the sum given by the solid lines. Even for moderately large values of $D$, this approximation is very accurate for $p>3$. 

It is possible to evaluate the sum within this approximation by taking the limit $D\xrightarrow{}\infty$ and converting it to an integral by defining the continuous parameter $x := \pi(n+1)/(D+1)$, which takes values in $[0,\pi)$:
\begin{align}\label{integral_apx}
    \begin{split}
        \lim_{D\xrightarrow{}\infty}\sum_{n=0}^{D-1}f^{(p,0)}_n = & -\frac{\pi^{7/2}p^2}{8D^4}\frac{\Gamma\left(\frac{2+p}{2}\right)}{\Gamma\left(\frac{1+p}{2}\right)} \\ & \cdot\int_0^\pi dx\left(1+\cot^2\left(x\right)\right)^2\sin^p\left(x\right).
    \end{split}
\end{align}
After evaluating the integral in the RHS of this equation, a formula for $\ell$ is obtained from Eq.~(\ref{g1_phase_diff_appx}) and, in turn, the coherence for the $p$-family,
\begin{align}\label{analytic_c0}
    \mathfrak{C}^{(p,0)} = \frac{256}{\pi^4p^2}\frac{\Gamma\left(\frac{p+1}{2}\right)\Gamma\left(\frac{p-2}{2}\right)}{\Gamma\left(\frac{p+2}{2}\right)\Gamma\left(\frac{p-3}{2}\right)}\mu^4, \quad \quad (D\xrightarrow{}\infty, p>3).
\end{align}

Eq.~(\ref{analytic_c0}) is compared with the numerical evaluation of $\mathfrak{C}$ for $D=1000$ in Fig.~\ref{fig:5}a. Excellent quantitative agreement is observed between these two methods of analysis for all values of $p\gtrsim 3$. Indeed, Eq.~(\ref{analytic_c0}) correctly predicts the peak in the coefficient for the power law $\mathfrak{C}=c\mu^4$, being at $p=4.1479$. This optimal value can be found by evaluating the stationary point for the prefactor in the above equation as a function of $p$. We begin to see significant deviations between our numerical and analytical results within the lower extreme for which Eq.~(\ref{analytic_c0}) is defined, where $3< p< 4$. We expect that numerical results for larger system sizes would yield better agreement with the analytical formula in this region, although showing this is exceedingly computationally expensive and therefore not addressed in the present Paper.

\subsection{Sub-Poissonian Laser Families}

Unfortunately, one cannot directly apply the arguments detailed above to obtain formulae for the coherence of the more general $p,\lambda$- and $p,q$-families, which exhibit sub-Poissonian beam photon statistics. The reason for this can be seen in Fig.~\ref{fig:8}b, which shows $g^{(1)}(\tau)$ for these two families for the choice of parameters that maximize $\mathfrak{C}$ and minimize $Q$, along with that produced by an ideal beam, $g^{(1)}(\tau)=\exp(-\ell|t|/2)$. Here, we can see that although the first order Glauber coherence functions of these families are essentially identical to the characteristic exponential decay exhibited by an ideal laser for relatively long times, small deviations between these two functions are observed for relatively short correlation times. This invalidates the first-order approximation given in Eq.~(\ref{g1_phase_diff_appx}).

In order to work around this, we take a heuristic approach and find that replacing Eq.~(\ref{g1_phase_diff_appx}) with the ansatz
\begin{align}\label{ansatz}
    \sum_{n=0}^{D-1}\left(\mathcal{L}\varrho^{\phi=0}_{\rm c}\right)_{n,n+1}\approx -\frac{\ell}{2},
\end{align}
corrects the small deviation seen in the short-time behaviour observed in Fig.~\ref{fig:8}b. Here we utilize the state $\varrho^{\phi}_{\rm c}$, which is a projector onto the pure cavity state
\begin{align}
    \ket{\psi^\phi}_c = \sum_{n=0}^{D-1}\sqrt{\rho_n}e^{i\phi n}\ket{n},
\end{align}
such that the uniformly weighted ensemble reproduces the incoherent steady-state of our families of laser models, i.e, $\int_0^{2\pi}\varrho_c^\phi d\phi/(2\pi) = \sum_{n=0}^{D-1}\rho_n\ket{n}\bra{n}$.
The LHS of Eq.~(\ref{ansatz}) can be thought as an average of the decay of the off-diagonal components of this pure cavity state, however we do not provide a rigorous justification for this equation. Regardless, it will be shown that this expression provides a remarkably powerful method of computing the coherence. In particular, by starting with this instead of Eq.~(\ref{g1_phase_diff_appx}) and following the steps which led to the general expression of the laser coherence for the $p$-family, similar expressions are able to be derived for the $p,\lambda$- and $p,q$-families, which are just as accurate.

With this caveat, we proceed to derive a general formula for the coherence of our $p,\lambda$- and $p,q$-families of laser models. Mirroring the steps outlined in Part A of this section, we have the sum elements $f_n^{(p,\lambda)} = \left(\mathcal{L}_{\rm M}^{(p,\lambda)}\varrho^{\phi=0}_c\right)_{n,n+1}$ and $f_n^{(p,q)} = \left(\mathcal{L}_{\rm NM}^{(p,q)}\varrho^{\phi=0}_c\right)_{n,n+1}$, which can be expressed as
\begin{subequations}\label{f_lamq_full}
\begin{align}
    \begin{split}\label{f_lamq_full1}
        f^{(p,\lambda)}_n = \frac{-\sqrt{\rho_{n}\rho_{n+1}}}{2}\Bigg{\{} {} & \left[ \left(\frac{\rho_{n-1}}{\rho_{n}} \right)^{\frac{(1-\lambda)}{2}} - \left( \frac{\rho_{n}}{\rho_{n+1}}\right)^{\frac{(1-\lambda)}{2}} \right]^2 \\ & + \left[ \left(\frac{\rho_{n+1}}{\rho_{n}} \right)^{\frac{\lambda}{2}} - \left( \frac{\rho_{n+2}}{\rho_{n+1}}\right)^{\frac{\lambda}{2}} \right]^2 \Bigg{\}},
    \end{split}
\end{align}
\begin{align}
    \begin{split}\label{f_lamq_full2}
        f^{(p,q)}_{n} & = \frac{-\sqrt{\rho_n\rho_{n+1}}}{2}\left[\left(\frac{\rho_{n-1}}{\rho_{n}}\right)^{\frac{(2+q)}{4}} - \left(\frac{\rho_{n}}{\rho_{n+1}}\right)^{\frac{(2+q)}{4}}\right]^2.
    \end{split}
\end{align}
\end{subequations}
Eq.~(\ref{f_lamq_full2}) has been obtained slightly less rigorously compared to Eq.~(\ref{f_lamq_full1}), which is largely due to the non-trivial dynamics imposed by the non-Markovian pumping process. Nonetheless, the resulting formula for $\mathfrak{C}$ derived through these means accurately matches our numerical evaluations. Performing a Taylor series expansion about the point $\pi/(D+1) = 0$ in Eqs.~(\ref{f_lamq_full1})~and~(\ref{f_lamq_full2}) gives, respectively, 
\begin{subequations}\label{f_lamq_apx}
\begin{align}
    f_n^{(p,\lambda)} \approx (2\lambda^2-2\lambda+1)f^{(p,0)}_n,
\end{align}
\begin{align}
    f_n^{(p,q)} \approx (1+q/2)^2f^{(p,0)}_n.
\end{align}
\end{subequations}
Eqs.~(\ref{f_lamq_full}) are compared with Eqs.~(\ref{f_lamq_apx}) in Fig.~\ref{fig:7}c-f. It can be seen there that the behaviour of the edge elements for these sums with respect to $p$ is the same as that for the $p$-family, and the approximations given in Eq.~(\ref{f_lamq_apx}) are very accurate. Applying the same arguments as those used to move from Eq.~(\ref{approx_elem}) to Eq.~(\ref{analytic_c0}) leads directly to an expression of the coherence for the $p,\lambda$- and $p,q$-families, respectively, with
\begin{subequations}
\begin{align}\label{coh_lam_general}
    \mathfrak{C}^{(p,\lambda)} = \frac{\mathfrak{C}^{(p,0)}}{2(\lambda-1/2)^2+1/2},
\end{align}
\begin{align}\label{coh_q_general}
    \mathfrak{C}^{(p,q)} = \frac{\mathfrak{C}^{(p,0)}}{(1+q/2)^2},
\end{align}
\end{subequations}
For $D\xrightarrow{}\infty, p>3$ in both cases. Eqs.~(\ref{coh_lam_general})~and~(\ref{coh_q_general}) correctly predict the coherence for the respective families of laser models for the same values of $p$ for which Eq.~(\ref{analytic_c0}) is valid. This agreement is also shown in Figs.~\ref{fig:5}c~and~\ref{fig:5}e, where the two equations given directly above are compared with iMPS calculations of the coherence for $D=1000$ for specific parameter choices that minimize the $Q$ parameter of the beam.

\section{Discussion and Conclusions}
\label{summary}

In this Paper, we have expanded upon the seminal result of Ref.~\cite{HL} by providing a study of lasers which exhibit Heisenberg-limited coherence as well as sub-Poissonian beam photon statistics. Much of this serves to supplement the companion Letter~\cite{Ostrowski}, which summarizes the key results of this study, while also providing many additional findings that substantially develop our understanding of laser models exhibiting Heisenberg-limited coherence. 

In particular, we have detailed the generalized proof of the upper bound for laser coherence, showing that $\mathfrak{C}=O(\mu^4)$ is the Heisenberg limit for a much broader class of lasers. This class of lasers encompass particular beams which can have a Mandel-$Q$ parameter for long photon counting durations on the beam arbitrarily close to the minimum of $Q=-1$~\cite{ZouMandel}. From this result, we introduced three new families of laser models; all of which were shown to demonstrate Heisenberg scaling of $\mathfrak{C}$ with $\mu$ under appropriate parameter choices. Two of these families, namely the $p,\lambda$-family and
$p,q$-family, which respectively exhibited a non-isometric, Markovian gain process (parameterized by $\lambda$) and a partially-isometric, non-Markovian gain process (parameterized by $q$), were shown to have sub-Poissonian beam photon statistics, with negative values of $Q$ under particular parameter values. For the $p,\lambda$-family, a minimum of $Q=-0.5$ was obtained, corresponding to a $50\%$ reduction of the photon noise in the beam below the shot-noise limit at cavity resonance. This was obtained for the choice of parameter $\lambda=0.5$, corresponding to a scenario where the matrix elements specifying gain and loss into and out of the laser device were defined as reciprocals to one another. For the $p,q$-family, complete photon noise reduction in the beam was acquired ($Q=-1$) when the pumping of excitations into the cavity was done so in a completely regular manner, corresponding to the choice of parameter $q\xrightarrow{}-1$.

This in turn led the to a central result of this Paper, and that of the companion Letter to this work~\cite{Ostrowski}, as it was found that the exact choice of parameters which minimize the $Q$ parameter of the beam (\ie, those which maximize the degree of sub-Poissonianity) within the two sub-Poissonian families of laser models also maximized the coherence. In particular, in the asymptotic limit $\mu\xrightarrow{}\infty$, it was found that $\lambda$ influences the coherence of the $p,\lambda$-family by modulating that of the $p$-family with a Lorentzian function centered at $\lambda=0.5$, with a peak of $2$, and a FWHM of $\Delta\lambda = 1$ (see Eq.~(\ref{coh_lam_general})). Furthermore, $q$ was found to influence the coherence of the $p,q$-family by modulating that of the $p$-family with the function $1/(1+q/2)^2$ (see Eq.~(\ref{coh_q_general})). This led to the conclusion that a trade-off between coherence and sub-Poissonianity in Heisenberg-limited laser models does not exist. To the contrary, taking measures to minimize $Q$ also gives rise an increase in $\mathfrak{C}$ for the specific laser models studied here. This trade-off is well-known to not exist in laser models with coherence at the SQL~\cite{Ralph2004}. The result at hand therefore marks a generalization of these results to laser models which have a vastly smaller (Heisenberg-limited) rate of phase diffusion.

Identifying why this ``win-win" relationship occurs between coherence and sub-Poissonianity in general appears as a subtle problem, because of the fact that the gain and loss processes within these Heisenberg-limited laser systems are highly nonlinear. However, there is a relatively straightforward interpretation for this behaviour within the $p,q$-family of models, where physical insight may be drawn from the results gathered from Section~\ref{general_p}. It is clear for all the models presented above, achieving Heisenberg-limited laser coherence requires the cavity number distribution to be much broader than a Poissonian distribution. However, as indicated in Fig.~\ref{fig:7}, there are constraints to how broad the optimal distribution should be. There, it can be seen that detrimental effects of the edge elements in the dynamics become present at relatively low values for $p$, while suggesting an optimal cavity distribution of $\rho_n \propto \sin^{4.15}(\pi(n+1)/(D+1))$. The effect of regularizing the pump within the $p,q$-family results in a reduction in the intensity noise introduced to the system via the pumping process; this will tend to narrow the intracavity distribution. In order to compensate this, and preserve the optimal cavity distribution, the elements of the loss operator, $L_{n}^{(p,-q/2)}$, defined in Eq.~(\ref{GainLoss}b) become ``flatter" as a function of $n$. Intuitively, this will decrease the phase noise added to the system by the loss process, since we have seen that both the gain and loss must be close to flat to achieve a Heisenberg-limited coherence. Hence, having a more regular pumping scenario will increase the coherence compared to a less regular one (higher values of $q$).

This physical interpretation is clearly model-specific, and does not explain the compatible relationship between coherence and sub-Poissonianity within the $p,\lambda$-family. Insight to this question more generally may be achieved through an analysis of these systems in different parameter regimes, where the gain and loss processes may be linearized allowing for analytical results for the first- and second-order Glauber coherence to be more easily obtained. These investigations will be a topic of future work.

Along with the aforementioned results, we were also able to derive formulae for the coherence of the three families of laser models which accurately reproduced the numerical results for large $\mu$. These formulae hold in the regime defined by $p\gtrsim3$, where the laser families exhibit Heisenberg-limited coherence, $\mathfrak{C}=\Theta(\mu^4)$, in the asymptotic limit $\mu\xrightarrow{}\infty$. In particular, these formulae support the notion that $p=4.1479$ is optimal to maximize the coherence in each family, which deviates slightly from the value $p=4$ used as an ansatz in Ref.~\cite{HL}. Although these formulae do not readily generalize to the regime of $p\lesssim3$, where the $\mu^4$ scaling of the coherence is lost, we were able to provide heuristic arguments based on our analysis to explain this change of scaling.

Looking outwards towards additional future work, a number of avenues are open. Firstly, as we have not provided any rigorous justification for the use of Eq.~(\ref{ansatz}), one could investigate why this ansatz serves as such an accurate formula to predict the linewidth. In particular, why does this equation work in general for the $p,\lambda$- and $p,q$-families (\ie, those which exhibit sub-Poissonian beam statistics), while a simple first order expansion in $t$ of Eq.~(\ref{g1_phase_diff}) does not? Secondly, one could also explore the fundamental limits for laser coherence under different assumptions on the beam. For example, we believe the limit found in Ref.~\cite{HL} can be tightened with a more elegant condition on the beam, being that it is exactly describable by a coherent state undergoing pure phase diffusion~\cite{Louisell,Carmichael}. This would be a much stricter requirement on the beam compared to Condition 4 of both this Paper and that in Ref.~\cite{HL}, as it would also place constraints on the Glauber coherence functions given in Eq.~(\ref{n_corrs_norm}) above second order. Preliminary numerical results suggest that this regime, where the beam may be described by a coherent state with diffusing phase, may be realised with our $p$-family in the limit $D\xrightarrow{}\infty$ and $p\gg1$.

\begin{acknowledgments}
The authors acknowledge the support of the Griffith University eResearch Service and Specialised Platforms Team and the use of the High Performance Computing Cluster ``Gowonda" to complete this research. This research was funded by the Australian Government through the through the Australian Research Council's Discovery Projects funding scheme (project DP220101602), and an Australian Government Research Training Program (RTP) Scholarship.
\end{acknowledgments}

\appendix

\section{Summary of Lemmas Required for the Proof of Theorem 1}
\label{lemmas}

In this appendix, we explicitly state each lemma required for the proof of the upper bound for $\mathfrak{C}$ given in the main text. Rigorous mathematical justifications for Lemmas 1--4 may be found in the SM of Ref.~\cite{HL}, while that of Lemma 5 may be found in Ref.~\cite{Bandilla1991}.

For the following, we let $\mathcal{B}(\mathcal{H})$ denote the set of bounded linear operators on the Hilbert space $\mathcal{H}$. Furthermore, the evolution of the system from time $T$ to $T'$ must be unitary, which can be generally described by the superoperator $\mathcal{U}_{\rm ce}^{T\xrightarrow{}T'}:\mathcal{B}(\mathcal{H}_{\rm c}\otimes\mathcal{H}_{\rm e})\xrightarrow{}\mathcal{B}(\mathcal{H}_{\rm c}\otimes\mathcal{H}_{\rm b}\otimes\mathcal{H}_{\rm e'})$. Here, $\mathcal{H}_{\rm c}$ and $\mathcal{H}_{\rm e}$ represent the Hilbert spaces associated with the laser device and initial environment states, respectively, and the subsequent primed environment, $\mathcal{H}_{\rm e'}$, contains everything not counted as part of the laser or beam (the latter of which being associated with the Hilbert space $\mathcal{H}_{\rm b}$) following the action of this superoperator. Following sHolevo~\cite{sHolevo}, we also use the term \textit{covariant} measurement to refer to any positive operator valued measure (POVM) element $\hat{E}^\phi$, with $\phi$ being its outcome, that obeys $\mathcal{U}^{\theta}(\hat{E}^\phi) = \hat{E}^{(\phi+\theta)}$, where the superoperator $\mathcal{U}^{\theta}(\bullet)$ describes a phase shift by arbitrary angle $\theta$.

\textbf{Lemma 1} (Laser encoded phase from filtering) \textit{Suppose at time $T$ the laser device is in a phase invariant state $\rho_{\rm c}^{\rm inv}$, and the laser, beam and environment are evolved up to time $T'$ by the unitary map $\mathcal{U}_{\rm ce}^{T\xrightarrow{}T'}:\mathcal{B}(\mathcal{H}_{\rm c}\otimes\mathcal{H}_{\rm e})\xrightarrow{}(\mathcal{H}_{\rm c}\otimes\mathcal{H}_{\rm b}\otimes\mathcal{H}_{\rm e'})$. If a covariant phase measurement is performed on the beam emitted over the interval $[T,T')$, and an outcome $\phi_F$ is obtained at time $T'$, the conditioned state of the laser is equivalent to a fiducial state $\rho^{\rm fid}_{\rm c}$ with an optical phase $\phi_F$ encoded by the generator $\hat{n}_{\rm c}$ (the number operator for excitations stored within the laser device). That is,
\begin{align}
    \rho_{\rm{c}|\phi_F}(T') = \mathcal{U}_{\rm c}^{\phi_F}(\rho^{\rm fid}_{\rm c}),
\end{align}
where $\mathcal{U}_{\rm c}^{\phi_F}(\rho_{\rm c}):=e^{i\phi_F\hat{n}_{\rm c}}\rho_{\rm c}e^{-i\phi_F\hat{n}_{\rm c}}$, and the fiducial state $\rho^{\rm fid}_{\rm c}$ is independent of $\phi_F$.}

\textbf{Lemma 2} (Steady state phase shift invariance). \textit{If there is a unique steady state of the laser, and Condition 2 (Endogenous Phase) holds, this steady state is invariant under all optical phase shifts,
\begin{align}
    \mathcal{U}^{\theta}_{\rm c}(\rho^{\rm ss}_{\rm c}) = \rho^{\rm ss}_{\rm c} \quad \forall \theta.
\end{align}}

\textbf{Lemma 3} (Phase encoding preserves photon statistics). \textit{For arbitrary covariant phase measurements on the beam, the fiducial state of the laser, $\rho_{\rm c}^{\rm fid}$, given in Lemma 1 has the same photon number statistics as the phase-invariant steady state, $\bra{n}\rho_{\rm c}^{\rm fid}\ket{n} = \bra{n}\rho_{\rm c}^{\rm ss}\ket{n}$.}

\textbf{Lemma 4} (Phase covariance). \textit{The filtering (retrofiltering) observable $e^{i\hat{\phi}_{F(R)}}$ changes covariantly when the beam undergoes an optical phase shift by arbitrary angle $\theta$. That is, the probability to obtain the result $e^{i\phi_{F(R)}}$ changes as
\begin{align}
    P(e^{i\phi_{F(R)}}|\mathcal{U}^\theta_{\rm b}(\rho_{\rm b})) = P(e^{i(\phi_{F(R)}-\theta)}|\rho_{\rm b}).
\end{align}}

\textbf{Lemma 5} (Minimum MSE for a phase measurement). \textit{An optical phase measurement on the state of a system with mean excitation number $\mu$, and for which the $U(1)$-mean phase is $\bar{\phi}$, will give an estimate $\hat{\phi}$ with MSE bounded from below by
\begin{align}\label{lemma5}
    1 - |\langle e^{i(\hat{\phi} - \bar{\phi})}\rangle|^2\geq4|z_A/3|^3\mu^{-2},
\end{align}
in the asymptotic limit $\mu\xrightarrow{}\infty$, where $z_A\approx-2.338$ is the first zero of the Airy function.}

\section{Mathematical Details for the $p,q$-family; Steady-State and Approximate master equation}
\label{RP_Appendix}

In this appendix we provide mathematical details relating to the family exhibiting the regularly pumped (non-Markovian),
quasi-isometric gain model (\ie, the $p,q$-family) described by Eq.~(\ref{mastereqq}) in the main text. In particular, we demonstrate that in the asymptotic limit, $D\xrightarrow{}\infty$, the cavity distribution at steady-state, $\rho_n = \bra{n}\rho_{\rm ss}\ket{n}$, for this family of models is characterised by the elements
\begin{align}\label{rp_dist}
    \rho_n = \alpha\sin^p\left(\pi\frac{n+1}{D+1}\right).
\end{align}
That is, it is the same as that for the other two families of laser models considered within this Paper. Additionally, we justify the various assumptions made throughout the derivation leading to the Master Equation~(\ref{mastereqq}) describing this family.

Addressing the cavity distribution first, the goal is to show that $\lim_{D\xrightarrow{}\infty}\mathcal{L}_{\rm NM}^{(p,q)}\rho_{ss} = 0$. From Eq.~(\ref{mastereqq}), we have (excluding edge elements, as they are of order $O(D^{-(p+1)})$ and therefore negligible)
\begin{align}
    \begin{split}
        \bra{n}\mathcal{L}_{\rm NM}^{(p,q)}\rho_{ss}\ket{n} = \bigg{\{} & (1-q)\frac{\rho_{n-1}}{\rho_{n}} + (q/2-1) \\ & + (q/2)\frac{\rho_{n-2}}{\rho_{n}} + \left(L^{(p,-q/2)}_{n+1}\right)^2\frac{\rho_{n+1}}{\rho_{n}} \\ & - \left(L^{(p,-q/2)}_{n}\right)^2 \bigg{\}}\rho_n.
    \end{split}
\end{align}
On substituting the expression for $L^{(p,-q/2)}_{n}$ given by Eq.~(\ref{GainLoss}b) in the main text with $x\xrightarrow{}-q/2\in(-\infty,1/2)$, we have
\begin{align}\label{ss_substituted}
    \begin{split}
        \bra{n}\mathcal{L}_{\rm NM}^{(p,q)}\rho_{ss}\ket{n} = \bigg{\{} & (1-q)\frac{\rho_{n-1}}{\rho_{n}} + (q/2-1) \\ & + (q/2)\frac{\rho_{n-2}}{\rho_{n}} + \left(\frac{\rho_{n}}{\rho_{n+1}}\right)^{q/2} \\ & - \left(\frac{\rho_{n-1}}{\rho_{n}}\right)^{(1+q/2)} \bigg{\}}\rho_n.
    \end{split}
\end{align}
It is possible to show using the elementary identity $\sin(A\pm B) = \sin(A)\cos(B)\pm\cos(A)\sin(B)$ that
\begin{align}\label{p/p}
    \frac{\rho_{n-1}}{\rho_n} = \left[\cos(\frac{\pi}{D+1}) - \sin(\frac{\pi}{D+1})\cot(\pi\frac{n+1}{D+1})\right]^p.
\end{align}
For sufficiently large $p$, the distribution $\rho_n$ is will be well-localized around its midpoint, $\mu$. In this scenario, the only relevant terms in Eq.~(\ref{ss_substituted}) will be those for which Eq.~(\ref{p/p}) may be treated as a linear function of $n$, that is
\begin{align}\label{apx_p/p}
    \frac{\rho_{n-1}}{\rho_n} = 1 + \frac{p\pi^2}{(D+1)^2}(n+1-\mu)+O((D+1)^{-3}),
\end{align}
where $n-\mu=O(1)$.
Substituting this expression into (\ref{ss_substituted}), it is straightforward to show that that all terms within the curly braces of lower order than $O((D+1)^{-3})$ vanish. Therefore, Eq.~(\ref{rp_dist}) rapidly converges to the actual steady-state distribution in the limit $D\xrightarrow{}\infty$ and $1\ll p\ll D$.

\begin{figure}[H]
\includegraphics[width=1.0\columnwidth]{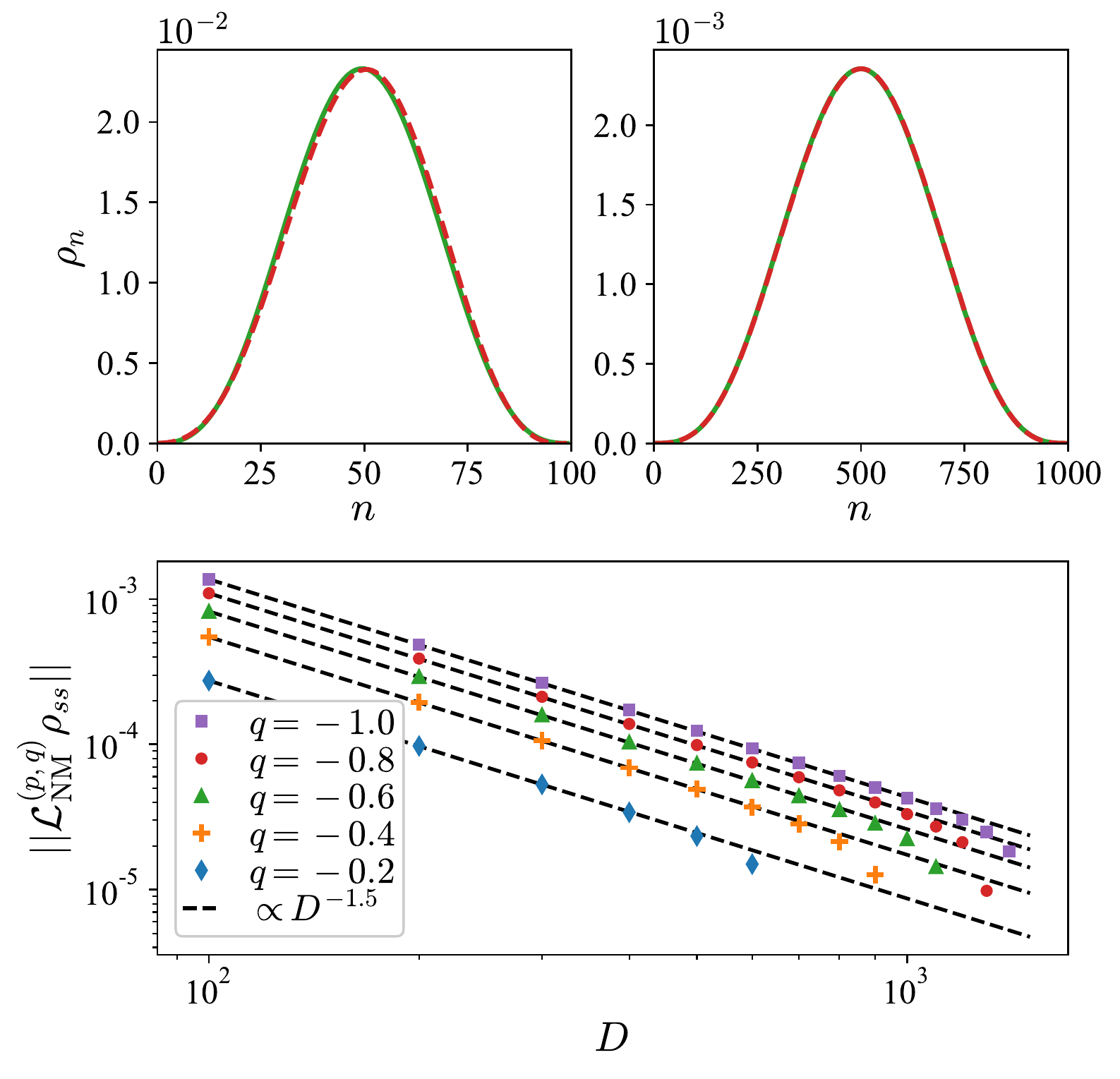}
\caption{\label{fig:A1} Top panels: Exact steady-state cavity distribution computed numerically from Eq.~(\ref{mastereqq}) (red dashed lines) compared with the distribution defined by Eq.~(\ref{rp_dist}) (solid green lines). Here, the choice of parameters are $p=3$ and $q=-1.0$. The top-left panel is for a relatively small cavity dimension ($D=100$), while the top-right panel is for $D=1000$. Bottom panel: Frobenius norms, defined as $||\hat{A}|| = (\rm{Tr}\{\hat{A}^\dagger\hat{A}\})^{-1/2}$, of the matrix $\mathcal{L}_{\rm NM}^{(p,q)}\rho_{ss}$ plotted against cavity dimension, $D$, for $p=3$ and a range of values of $q$. Dashed black lines correspond to a power law with an exponent of $-1.5$ as a guide for the eye.
}
\end{figure}

To illustrate this point, we compare the distribution defined by Eq.~(\ref{rp_dist}) to the exact steady-state of Eq.~(\ref{mastereqq}) computed numerically in the top panels of Fig.~\ref{fig:A1}. Here, we have chosen $q = -1$ and $p=3$. Such a value of $p$ is the smallest that we consider for our calculations of the coherence in Section~\ref{general_p} and is much smaller than what would be required for the linearized treatment of the loss operators given in Eq.~(\ref{apx_p/p}). Regardless of this, we still find Eq.~(\ref{rp_dist}) to be an excellent approximation of the actual steady-state distribution for the $p,q$-family. Even for moderately large values for the cavity dimension, such as that given in the right-hand top panel of Fig.~\ref{fig:A1} where $D=1000$, we find the two distributions to be virtually indistinguishable.

\begin{figure}[H]
\includegraphics[width=1.0\columnwidth]{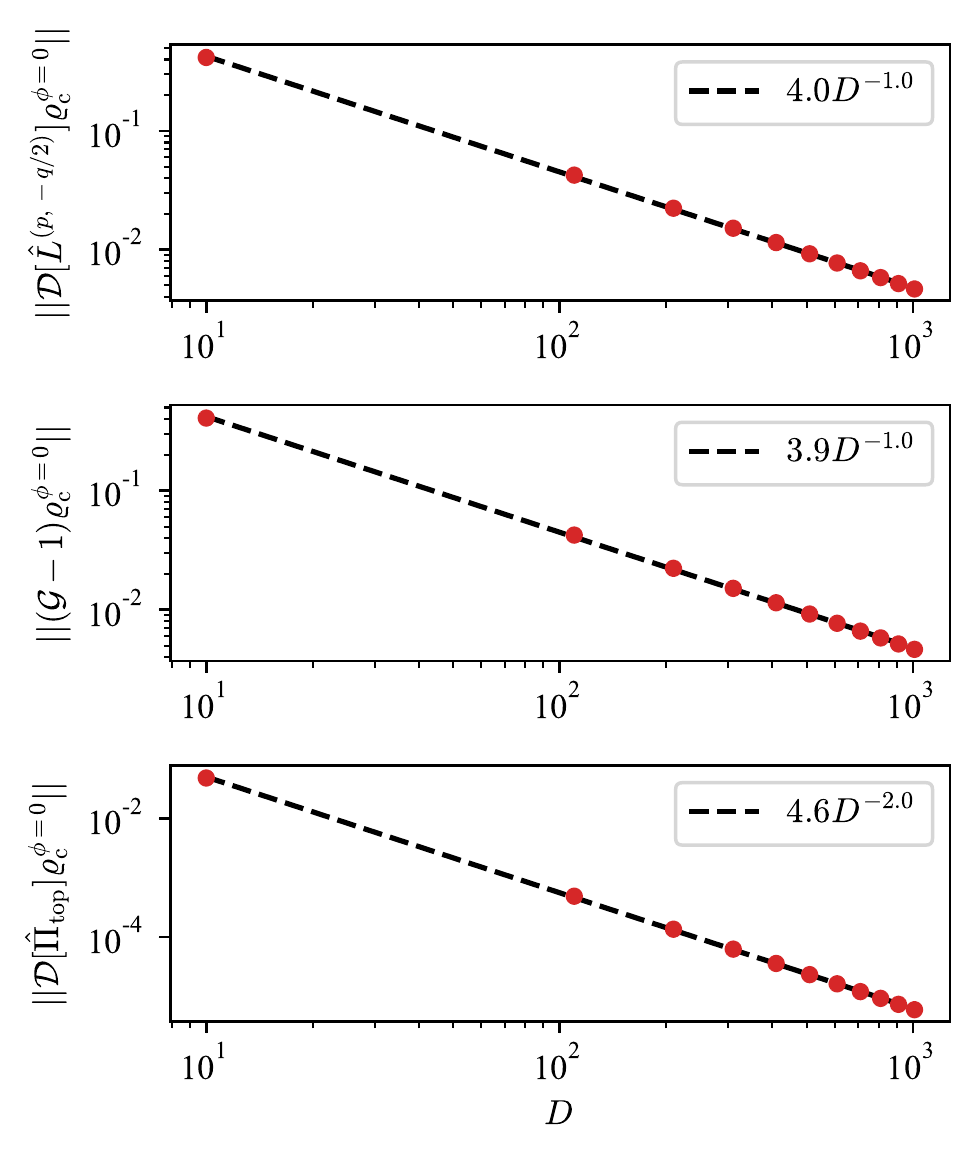}
\caption{\label{fig:A2} Frobenius norms, defined as $||\hat{A}|| = (\rm{Tr}\{\hat{A}^\dagger\hat{A}\})^{-1/2}$, of the matrices $\mathcal{D}[\hat{L}^{(p,-q/2)}]\varrho_{\rm c}^{\phi=0}$, $(\mathcal{G}-1)\varrho_{\rm c}^{\phi=0}$ and $\mathcal{D}[\hat{\Pi}_{\rm top}]\varrho_{\rm c}^{\phi=0}$ (red dots) as a function of cavity dimension, $D$. $\varrho_{\rm c}^{\phi=0}$ represents the pure cavity state, defined in Eq.~(\ref{p_cav_state}). Dashed black lines correspond to fitted power laws to this data.
}
\end{figure}

In order to be more rigorous and demonstrate the validity of Eq.~(\ref{rp_dist}) for more general values of $p$, we revert to numerical methods. In the bottom panel of Fig.~\ref{fig:A1}, we show the Frobenius norms of $\mathcal{L}_{\rm NM}^{(p,q)}\rho_{ss}$ against cavity dimension, $D$, for $p=3$ (again, which is the smallest value of $p$ for which our formula for the coherence, given by Eq.~(\ref{coh_q_general}), holds) and a range of values for $q$. Here, it is clear that for all values of $q$, this quantity converges to zero in the limit $D\xrightarrow{}\infty$, as each of the matrix norms are $o(D^{-1.5})$.

Next, we justify the various approximations made in the steps of the derivation which led to Eq.~(\ref{mastereqq}), the master equation for the $p,q$-family of laser models. In that derivation, it was assumed that over some short time interval, $\Delta t$, the superoperators giving rise to the gain and loss of excitations within the cavity (see Eqs.~(\ref{gain_loss_superops}) in the main text) act independently. For this to be the case, the action of the superoperators $\mathcal{D}[\hat{L}^{(p,-q/2)}]$ and $(\mathcal{G}-1) = \mathcal{D}[\hat{G}^{(p,0)}] + \mathcal{D}[\hat{\Pi}_{\rm top}]$ on the cavity state need to be small, in some sense. To demonstrate this, we consider their action on the pure cavity state
\begin{align}\label{p_cav_state}
    \varrho_{\rm c}^\phi = \sum_{n,m = 0}^{D-1}\sqrt{\rho_n\rho_m}e^{i\phi(n-m)}\ket{n}\bra{m}.
\end{align}
This state was also briefly introduced in Section~\ref{general_p} of the main text to derive formulae for the coherence of our families of laser models; it is defined in such a way that a uniformly weighted mixture reproduces the cavity steady-state, $\int_0^{2\pi}\varrho_c^\phi d\phi/(2\pi) = \sum_{n=0}^{D-1}\rho_n\ket{n}\bra{n}$. To motivate this choice of state, we assume that throughout the evolution of the system, the cavity state can be expressed as a mixture of states within the ensemble $\{\varrho_{\rm c}^\phi,d\phi/(2\pi)\}$; or in other words, we assume that this ensemble is physically realizable~\cite{Vaccaro2001}. A rigorous investigation of the validity of this pure state assumption will be addressed in a future Paper.

In Fig.~\ref{fig:A2} we show the Frobenius norms of the matrices $\mathcal{D}[\hat{L}^{(p,-q/2)}]\varrho_{\rm c}^{\phi=0}$ and $(\mathcal{G}-1)\varrho_{\rm c}^{\phi=0}$ against cavity dimension, $D$, with $q=-1$ and $p=3$. Fitting a power law to these points indicate that these Frobenius norms are $O(D^{-1})$. This implies that over the time interval, $\Delta t$, one may treat the gain and loss processes independently if one takes $\mathcal{N}\Delta t\ll D$, in which case
\begin{align}
    \begin{split}
        \rho(t+\Delta t) \approx & \left(1 + (\mathcal{G}-1)\right)^{n(\Delta t)}\rho(t) \\ & + \mathcal{N}\Delta t\mathcal{D}[\hat{L}^{(p,-q/2)}]\rho(t).
    \end{split}
\end{align}
Recall that we let $n(\Delta t) = \mathcal{N}\Delta t + \sqrt{\mathcal{N}(q+1)}\Delta W$ represent the number of excitations gained by the cavity in the interval $\Delta t$. Additionally, since we have $(\mathcal{G}-1)\varrho_{\rm c}^{\phi=0}=O(D^{-1})$, a binomial expansion to second order in $(\mathcal{G}-1)$ is reasonable, therefore validating Eq.~(\ref{binom_exp}) and allows one to write
\begin{align}
    \begin{split}
        \rho(t+\Delta t)\approx & \bigg{[}1 + (\mathcal{N}\Delta t + \sqrt{\mathcal{N}(q+1)}\Delta W)(\mathcal{G}-1) \\ & + \frac{1}{2}\big{\{}(\mathcal{N}\Delta t)^2 + 2\mathcal{N}\sqrt{\mathcal{N}(q+1)}\Delta t\Delta W \\ & + \mathcal{N}(q+1)(\Delta W)^2 - \sqrt{N(q+1)}\Delta W \\ & - \mathcal{N}\Delta t\big{\}}(\mathcal{G}-1)^2 + \mathcal{N}\Delta t\mathcal{D}[\hat{L}^{(p,-q/2)}]\bigg{]}\rho(t).
    \end{split}
\end{align}

Averaging over the uncertainty in the number of excitations added to the cavity and taking the limit $\Delta t \xrightarrow{} 0^+$, one obtains the master equation
\begin{align}\label{mastereqq_full}
    \begin{split}
        \frac{d\rho}{dt} & = \mathcal{N}\left((\mathcal{G}-1) +\frac{q}{2}(\mathcal{G}-1)^2 + \mathcal{D}[\hat{L}^{(p,q)}]\right)\rho.
    \end{split}
\end{align}
To obtain Eq.~(\ref{mastereqq}) in the main text, we also let $(\mathcal{G}-1)\approx\mathcal{D}[\hat{G}^{(p,0)}]$. This is also justified on inspection of Fig.~\ref{fig:A2}, where it is also shown that the Frobenius norm of $\mathcal{D}[\hat{\Pi}_{\rm top}]\varrho_{\rm c}^\phi$ decays much faster than that of $(\mathcal{G}-1)\varrho_{\rm c}^\phi$.

\section{Verification of Condition 4 for the Three Families of Laser Models}
\label{verification}

In this appendix, we verify the claim in Section~\ref{Numerical_Analysis} of the main text, that the families of laser models we introduced exhibit Heisenberg-limited beam coherence for certain parameter values. In that section, it was shown that the coherence of the beam scales as $\mu^4$ for the $p$-, $p,\lambda$- and $p,q$-families for values $p\gtrapprox3$. To say that this scaling is at the ultimate limit imposed by quantum mechanics, then Condition 4 must be verified, which means that Theorem 1 applies to these families of laser models. We remind the reader that for Condition 4 to be satisfied by a particular laser model, Eqs.~(\ref{c4.1})~and~(\ref{c4.2}) must hold. 

These two equations were already shown to hold for the $p$-family in Ref.~\cite{HL}, therefore we consider only the $p,\lambda$- and $p,q$-families here, which are those that can exhibit sub-Poissonian beam photon statistics. Moreover, we constrain the parameters for each of these families to $p=4.1479$, $\lambda=0.5$ and $p= 4.1479$, $q=-1$, respectively. These parameter values yield maximal beam coherence as well as a maximal degree of beam sub-Poissonianity (that is, a minimized value of $Q$) for each family. Therefore, it is expected that the deviations of the first- and second-order Glauber coherence functions between these families of laser models and that of an `ideal' beam would be at their greatest for these parameter values.

\begin{figure}[H]
\includegraphics[width=1.0\columnwidth]{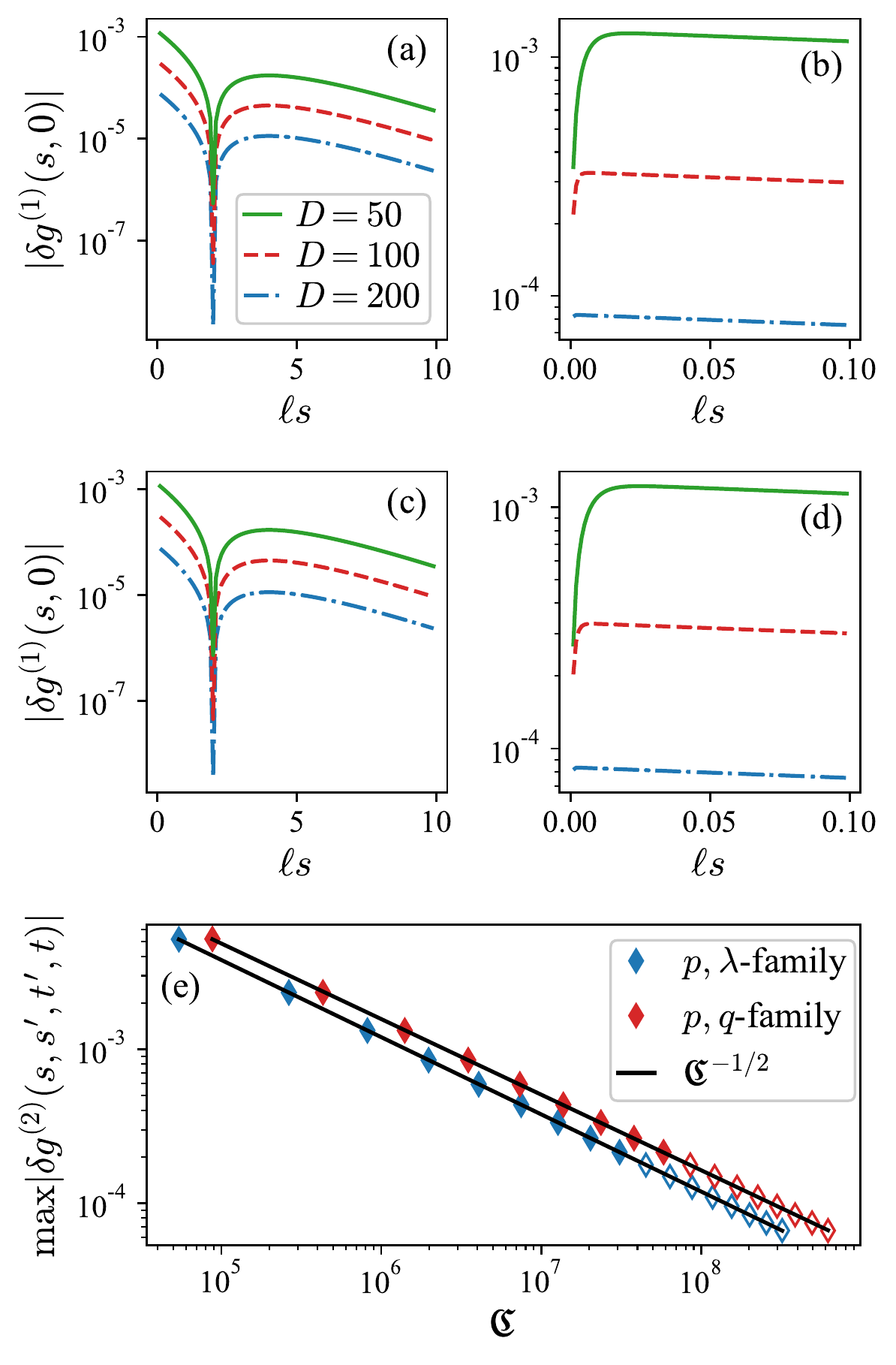}
\caption{\label{fig:B} (a): Deviations from the ideal laser model (a coherent state undergoing pure phase diffusion) of the first-order Glauber coherence function for the family of laser models exhibiting a randomly-pumped (Markovian), non-isometric gain ($p,\lambda$-family) over ten coherence times. solid green, red dashed and blue dash-dotted lines correspond to cavity dimensions $D=50,100,200$, respectively. (b): Same as that shown in (a), but over a much shorter timescale. (c,d): Same as that shown in (a,b), respectively, but for the family of laser models exhibiting a regularly-pumped (non-Markovian), quasi-isometric gain ($p,q$-family). (e): Global maxima of $|\delta g^{(2)}(\tau,s',t',t)|$ versus coherence (solid diamonds) calculated for $\{s',t',t\}\in[-\tau,\tau]$ employing interior-point optimizations of the iMPS forms for bond dimensions up to 250. Some examples of $|\delta g^{(2)}|$ are shown for bond dimensions up to $D=450$ (hollow diamonds). Blue and red correspond to the $p,\lambda$-family and $p,q$-family, respectively. Black solid lines correspond to power-law fits to the data, $|\delta g^{(2)}| = 1.2\mathfrak{C}^{-0.5}$ and $|\delta g^{(2)}| = 1.4\mathfrak{C}^{-0.5}$. For each of these families in plots (a-e), parameters are chosen such that beam coherence, $\mathfrak{C}$, is maximized for a given value of $D$, while $Q$ is also minimized.
}
\end{figure}

In order to demonstrate that Eqs.~(\ref{c4.1})~and~(\ref{c4.2}) hold for the $p,\lambda$- and $p,q$-families, we employ the same methods as those outlined in the supplementary material of Ref.~\cite{HL}. With regard to Eq.~(\ref{c4.1}), which places a condition on the first-order Glauber coherence function, time-translation invariance permits a direct search for the maximum deviation over only a single time argument. In Fig.~\ref{fig:B}a~and~\ref{fig:B}b, we show the quantity $|\delta g^{(1)}(s,0)|$, as defined by Eq.~(\ref{delta_gn}), for the $p,\lambda$-family over two different time scales. The same is also shown in Fig.~\ref{fig:B}c~and~\ref{fig:B}d for the $p,q$-family. As made clear from these plots, the largest deviations in the first-order coherence functions occur for each family at a short, non-zero time delay. Because this is decreasing as the cavity dimension, $D$, is increased, we may conclude that Eq.~(\ref{c4.1}) holds for the $p,\lambda$- and $p,q$-family for parameter values which maximize $\mathfrak{C}$ and minimize $Q$.

Verifying that Eq.~(\ref{c4.2}) also holds for our two sub-Poissonian families of laser models required more sophisticated numerical techniques. This is because the problem requires an optimization over three time parameters (note that one of the four parameters in Eq.~(\ref{c4.2}) may be removed from this optimization by imposing time translation invariance). This optimization was carried out by employing a highly-scalable, non-linear optimisation routine known as the interior point method described in Refs.~\cite{Int_point_1,Int_point_2,Int_point_3} to maximize Eq.~(\ref{G2mps}), along with all other non-trivial permutations of the bosonic operators~\cite{HL}. As required by Condition 4, the difference between any two of the time arguments was constrained to be $O(\sqrt{\mathfrak{C}}/\mathcal{N})$.

From each optimization, it was found that the time arguments which maximized $|\delta g^{(2)}(s,s',t',t)|$ were those for which there was negligible delay, for example $(s,s',t',t)=(s,s+\epsilon,s+2\epsilon,s+3\epsilon)$ with $\epsilon\xrightarrow{}0^+$. In Fig.~\ref{fig:B}e we plot ${\rm max}|\delta g^{(2)}(s,s',t',t)|$, where $s,s',t',t\in[-\tau,\tau]$ and $\tau = \sqrt{3/2\mathcal{N}\ell}$, for the $p,\lambda$- and $p,q$-families, respectively with blue and red diamonds. In particular, the results from the optimizations are indicated by the solid diamonds, which were performed for cavity dimensions ranging between $50\leq D\leq 250$. The hollow diamonds are extrapolations, where $|\delta g^{(2)}(s,s',t',t)|$ is computed for larger cavity dimensions using the same time arguments which were obtained from the optimizations. These deviations in $g^{(2)}$ are seen to scale as $\mathfrak{C}^{-1/2}$, therefore allowing us to conclude that Eq.~(\ref{c4.2}) also holds for the $p,\lambda$- and $p,q$-family of laser models and hence Condition 4 is satisfied. This allows the statement to be made that all three families of laser models considered in this Paper exhibit a beam coherence which is Heisenberg-limited.

\end{document}